\documentclass{jkas}

\def\year{} 
\def\month{} 
\def\volume{00} 
\def\issue{0} 
\def\beginpage{1} 
\def\endpage{99} 
\def\received{} 
\def\accepted{} 
\date{Received \received; accepted \accepted}

\title{
Decay of Turbulence in Fluids with Polytropic Equations of State
}

\author[1]{Jeonghoon Lim}
\author[1,2]{Jungyeon Cho}

\affil[1]{Department of Astronomy and Space Science, Chungnam National University, 99, Daehak-ro, Yuseong-gu, Daejeon, 34134, Republic of Korea; \email{jhlim0918@o.cnu.ac.kr}}
\affil[2]{Korea Astronomy and Space Science Institute, 776, Daedeokdae-ro, Yuseong-gu, Daejeon, 34055, Republic of Korea; \email{jcho@cnu.ac.kr}}

\def\corrauthor{
Jungyeon Cho
}

\def\runningauthor{
Jeonghoon Lim
}

\def\runningtitle{
Decay of Polytropic Turbulence
}

\def\keywords{
ISM: general - hydrodynamics - turbulence
}

\def\abstracttext{
We present numerical simulations of decaying hydrodynamic turbulence initially driven by solenoidal (divergence-free) and compressive (curl-free) driving. Most previous numerical studies for decaying turbulence assume an isothermal equation of state (EOS). Here we use a polytropic EOS, $P$ $\propto$ $\rho^{\gamma}$, with polytropic $\gamma$ ranging from 0.7 to 5/3. We mainly aim at determining the effects of polytropic $\gamma$ and driving schemes on the decay law of turbulence energy, E $\propto$ $t^{-\alpha}$. We additionally study  probability density function (PDF) of gas density and skewness of the distribution in polytropic turbulence driven by compressive driving. Our findings are as follows. First of all, we find that even if polytropic $\gamma$ does not strongly change scaling relation of the decay law, the driving schemes weakly change the relation; in our all simulations, turbulence decays with $\alpha$ $\approx$ 1, but compressive driving yields smaller $\alpha$ than solenoidal driving at the same sonic Mach number. Second, we calculate compressive and solenoidal velocity components separately and compare their decay rates in turbulence initially driven by compressive driving. We find that the former decays much faster so that it ends up having a smaller fraction than the latter. Third, the density PDF of compressively driven turbulence with polytropic $\gamma$ $>$ 1 deviates from log-normal distribution: it has a power-law tail at low density as in the case of solenoidally driven turbulence. However, as it decays, the density PDF becomes approximately log-normal. We discuss why decay rates of compressive and solenoidal velocity components are different in compressively driven turbulence and astrophysical implication of our findings.
}
\begin{document}
\jkashead 

\begin{table*}[ht!]
\centering
\caption{Simulation conditions\label{tab:tab1}}
\begin{tabular}{cccccc}
\toprule
 Run & Driving$^{\rm a}$ & $\gamma^{\rm b}$ & Resolution & $k_{ej}^{\rm c}$ & $M_{s}^{\rm d}$ \\
\midrule
SMS1-$\gamma$0.7  & Solenoidal & 0.7 & $512^3$ & 8.0 & $\sim$ 1  \\ 
SMS1-$\gamma$1.0  & Solenoidal & 1.0 & $512^3$ & 8.0 & $\sim$ 1  \\
SMS1-$\gamma$5/3  & Solenoidal & 5/3 & $512^3$ & 8.0 & $\sim$ 1  \\ \addlinespace
\midrule
SMS3-$\gamma$0.7  & Solenoidal & 0.7 & $512^3$ & 8.0 & $\sim$ 3  \\
SMS3-$\gamma$1.0  & Solenoidal & 1.0 & $512^3$ & 8.0 & $\sim$ 3  \\
SMS3-$\gamma$5/3  & Solenoidal & 5/3 & $512^3$ & 8.0 & $\sim$ 3  \\ \addlinespace
\midrule
SMS5-$\gamma$0.7  & Solenoidal & 0.7 & $512^3$ & 8.0 & $\sim$ 5  \\
SMS5-$\gamma$1.0  & Solenoidal & 1.0 & $512^3$ & 8.0 & $\sim$ 5  \\
SMS5-$\gamma$1.5  & Solenoidal & 1.5 & $512^3$ & 8.0 & $\sim$ 5  \\ \addlinespace
\midrule
CMS1-$\gamma$0.7  & Compressive & 0.7 & $512^3$ & 8.0 & $\sim$ 1 \\
CMS1-$\gamma$1.0  & Compressive & 1.0 & $512^3$ & 8.0 & $\sim$ 1 \\
CMS1-$\gamma$5/3  & Compressive & 5/3 & $512^3$ & 8.0 & $\sim$ 1 \\ \addlinespace
\midrule
CMS3-$\gamma$0.7  & Compressive & 0.7 & $512^3$ & 8.0 & $\sim$ 3 \\
CMS3-$\gamma$1.0  & Compressive & 1.0 & $512^3$ & 8.0 & $\sim$ 3 \\
CMS3-$\gamma$1.5  & Compressive & 1.5 & $512^3$ & 6.0 & $\sim$ 3 \\ \addlinespace
\midrule
CMS5-$\gamma$0.7  & Compressive & 0.7 & $512^3$ & 8.0 & $\sim$ 5 \\
CMS5-$\gamma$1.0  & Compressive & 1.0 & $512^3$ & 8.0 & $\sim$ 5 \\
\bottomrule
\end{tabular}
\tabnote{
$^{\rm a}$  Driving schemes - either solenoidal or compressive driving.    \\
$^{\rm b}$  Polytropic exponent.   \\
$^{\rm c}$  The driving wavenumber at which the energy injection rate peaks.   \\
$^{\rm d}$  The sonic Mach number which is defined in Equation (\ref{eq:eq5}).} \\
\end{table*}

\section{Introduction\label{sec:intro}}
Supersonic turbulence in the interstellar medium (ISM) is a well-known phenomenon and plays an essential role in star formation processes \citep{L1981,PN02,MK04}. Given that driving mechanisms of astrophysical turbulence are usually intermittent in both space and time, it is natural for turbulence to decay. Earlier studies showed that non-driven turbulence decays quickly in approximately one large-eddy turnover time (for hydrodynamic turbulence, see e.g., \citealt{Lesieur}; for magnetohydrodynamic turbulence, see \citealt{Mac98, Stone1998}), which is consistent with the fact that energy cascade occurs within one large-scale eddy turnover time even in the case of strongly magnetized turbulence \citep{GS95}.

It has been analytically suggested that turbulence energy decays with a power-law form of E $\propto$ t$^{-\alpha}$ (see e.g., chap.7 of \citealt{Lesieur}). Results from previous numerical studies of turbulence have converged that the value of $\alpha$ is approximately unity, and it does not strongly depend on the degree of magnetization and compressibility \citep{Mac98,Stone1998, BM99, Os01, CLV02}. 

Even if the consensus that turbulence quickly decays has been numerically established for the last two decades, the previous numerical results depend heavily on isothermal condition.
However, as long as various density and temperature phases in the ISM \citep{Fe01} are concerned, the use of polytropic equation of state (EOS)
\begin{equation}
\label{eq:eq1}
P=K\rho^{\gamma},
\end{equation}
where $P$ is the pressure, $\rho$ is the density, and both K and $\gamma$ are constants, is a valid approach (see \citealt{VPP96} and reference therein). The polytropic EOS has been used for many astrophysical problems, such as complex chemical processes \citep{SS00, GM07a}, or turbulence \citep{Scalo98,Li03,GM07b,F15}.

Besides a variety of density and temperature phases, a wide range of driving agents of turbulence also characterizes interstellar turbulence (see \citealt{F17} for a review). Based on its compressibility, we may consider two extreme types of driving; solenoidal (divergence-free) and compressive driving (curl-free). Until recently, solenoidal driving had been mainly used for turbulence studies. However, \citet{Fe10} showed that compressive driving and solenoidal driving can have different statistics. For example, they showed that ``the former yields stronger compression at the same RMS Mach number than the latter, resulting in a three times larger standard deviation of volumetric and column density probability distributions."  To our best knowledge, scaling relations of decaying polytropic turbulence initially driven by compressive driving have not been studied yet.

The main goal of this paper is to examine whether decay exponent $\alpha$ depends on the value of the polytropic $\gamma$. Here we concentrate on decay of polytropic turbulence  driven by either solenoidal or compressive driving in both transonic and supersonic regimes. Hence, we expect to demonstrate what impacts polytropic EOS and types of driving have on decaying turbulence. In addition, we also investigate probability density function (PDF) of gas density and skewness of the PDF in decaying polytropic turbulence initially driven by compressive driving.

The paper is organized as follows. We explain our motivation and numerical method in Section \ref{sec:sec2}, and present the results from our numerical simulations in Section \ref{sec:sec3}. We discuss our finding and its astrophysical implication and give summary in Section \ref{sec:sec4}.

\begin{figure*}
\centering
\includegraphics[scale=0.35]{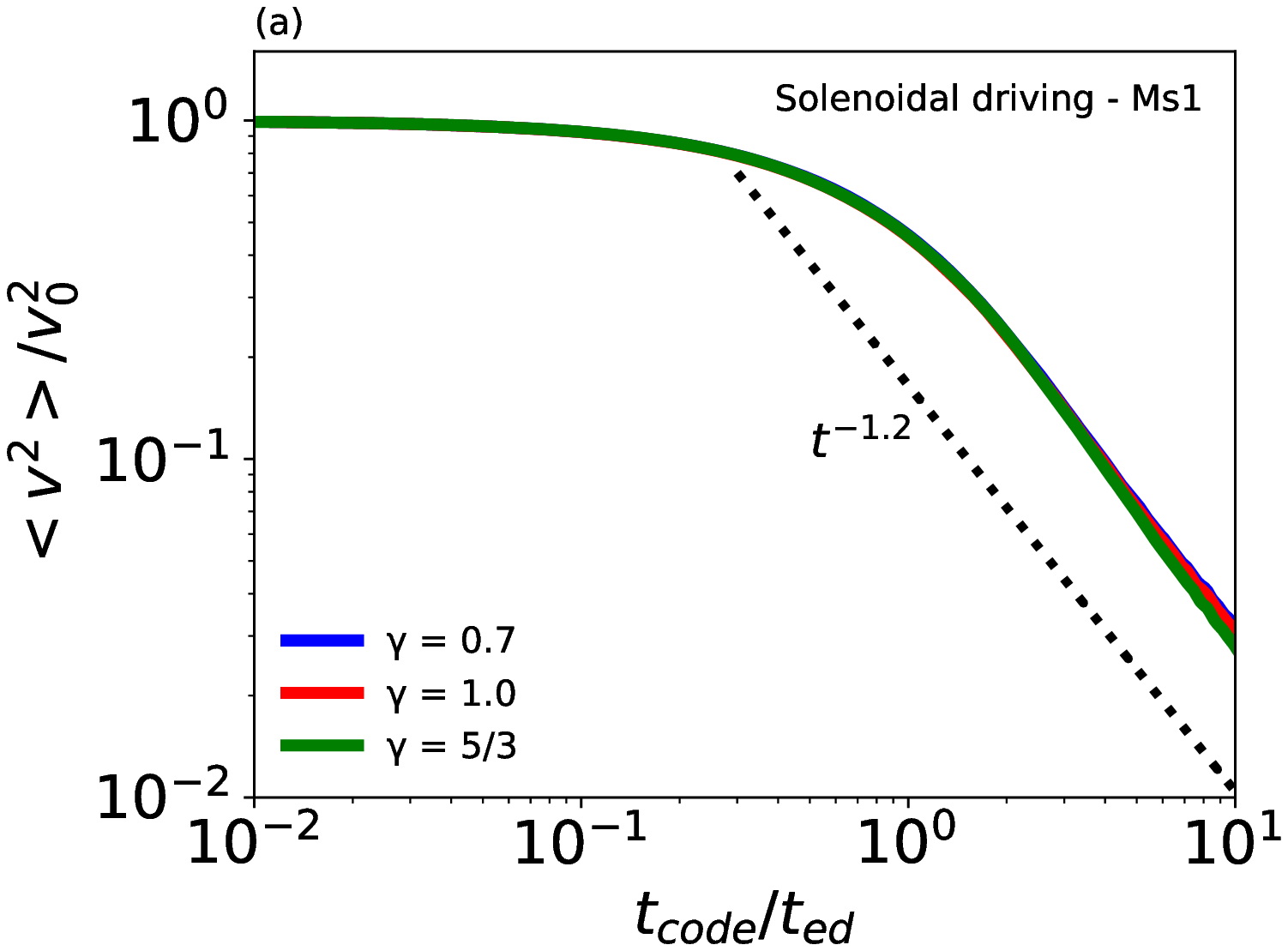}
\includegraphics[scale=0.35]{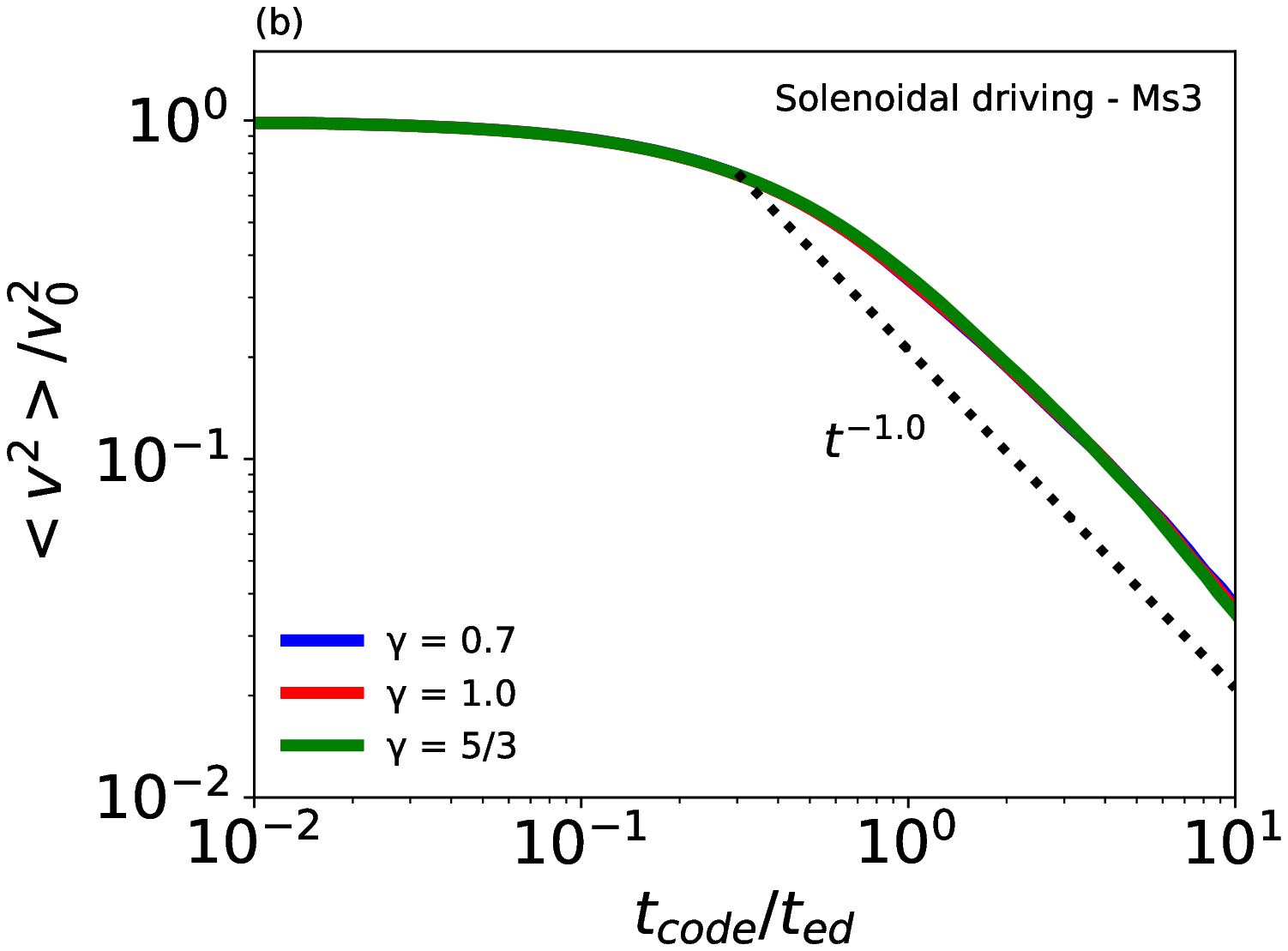}
\includegraphics[scale=0.35]{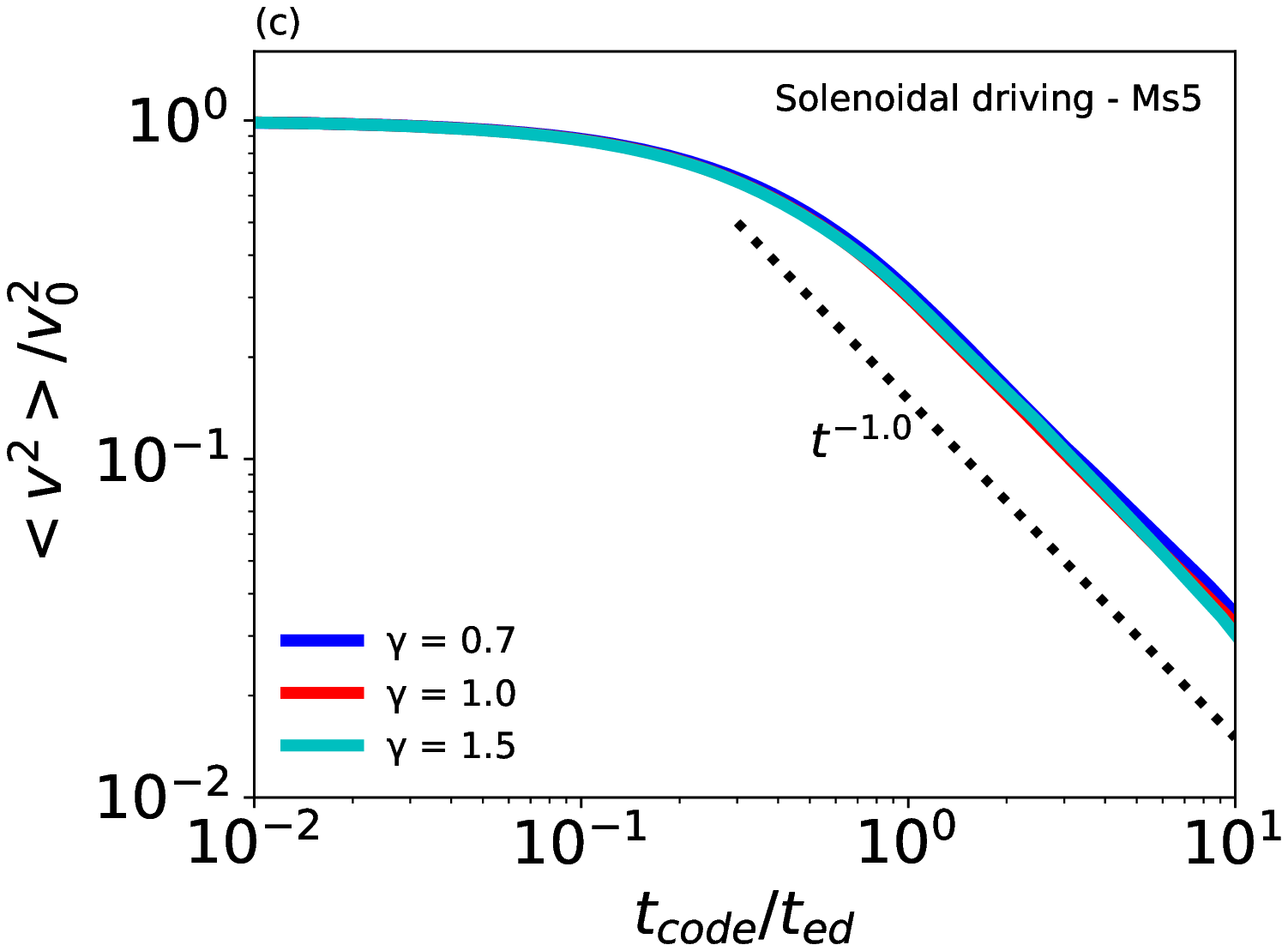}
\caption{Time evolution of spatial averaged kinetic energy density $<v^2>$ in decaying turbulence initially driven by solenoidal driving with polytropic $\gamma$ = 0.7 (blue), 1.0 (red), 1.5 (cyan), or 5/3 (green). Left panel: $Ms$ $\sim$ 1. Middle panel: $M_s$ $\sim$ 3. Right panel: $M_s$ $\sim$ 5. Turbulence starts decaying at $t_{code}/t_{ed}$ = 0. We normalize $<v^2>$ by $v_{0}^2$ which is the value at $t_{code}/t_{ed}$ = 0. The black dotted lines in all three panels are reference lines for different power-law exponents.
\label{fig:fig1}}
\end{figure*}


\begin{figure*}
\centering
\includegraphics[scale=0.35]{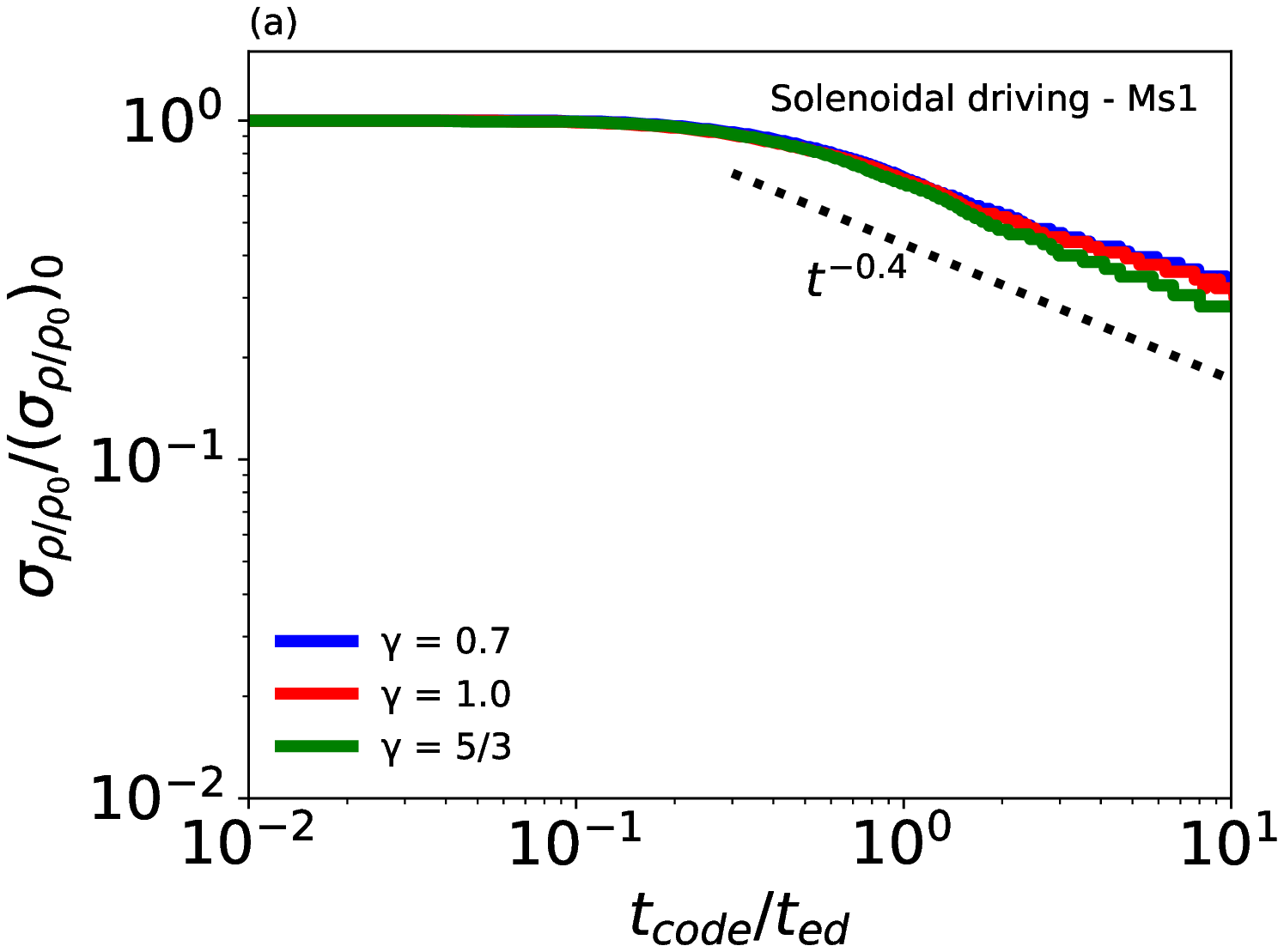}
\includegraphics[scale=0.35]{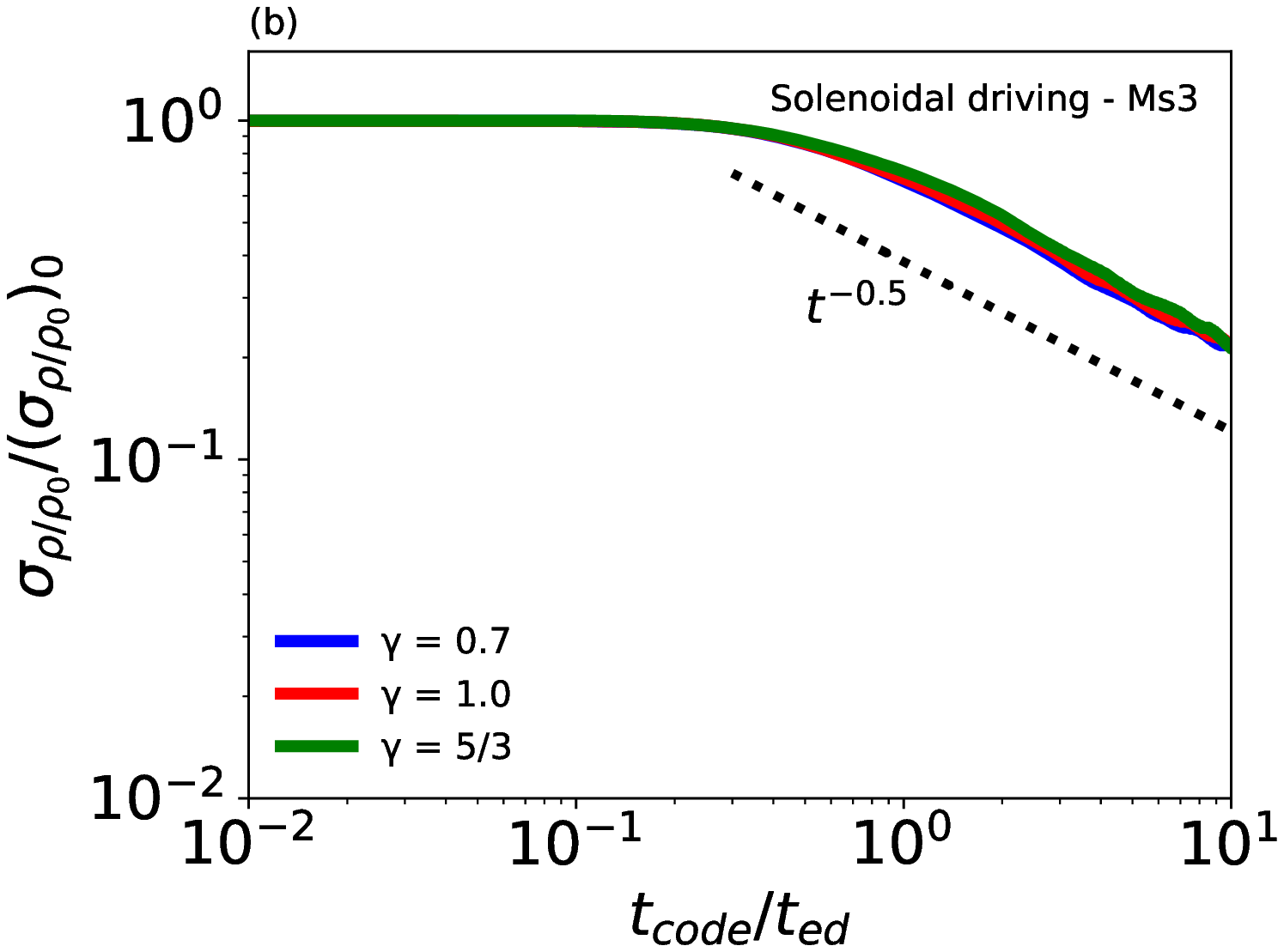}
\includegraphics[scale=0.35]{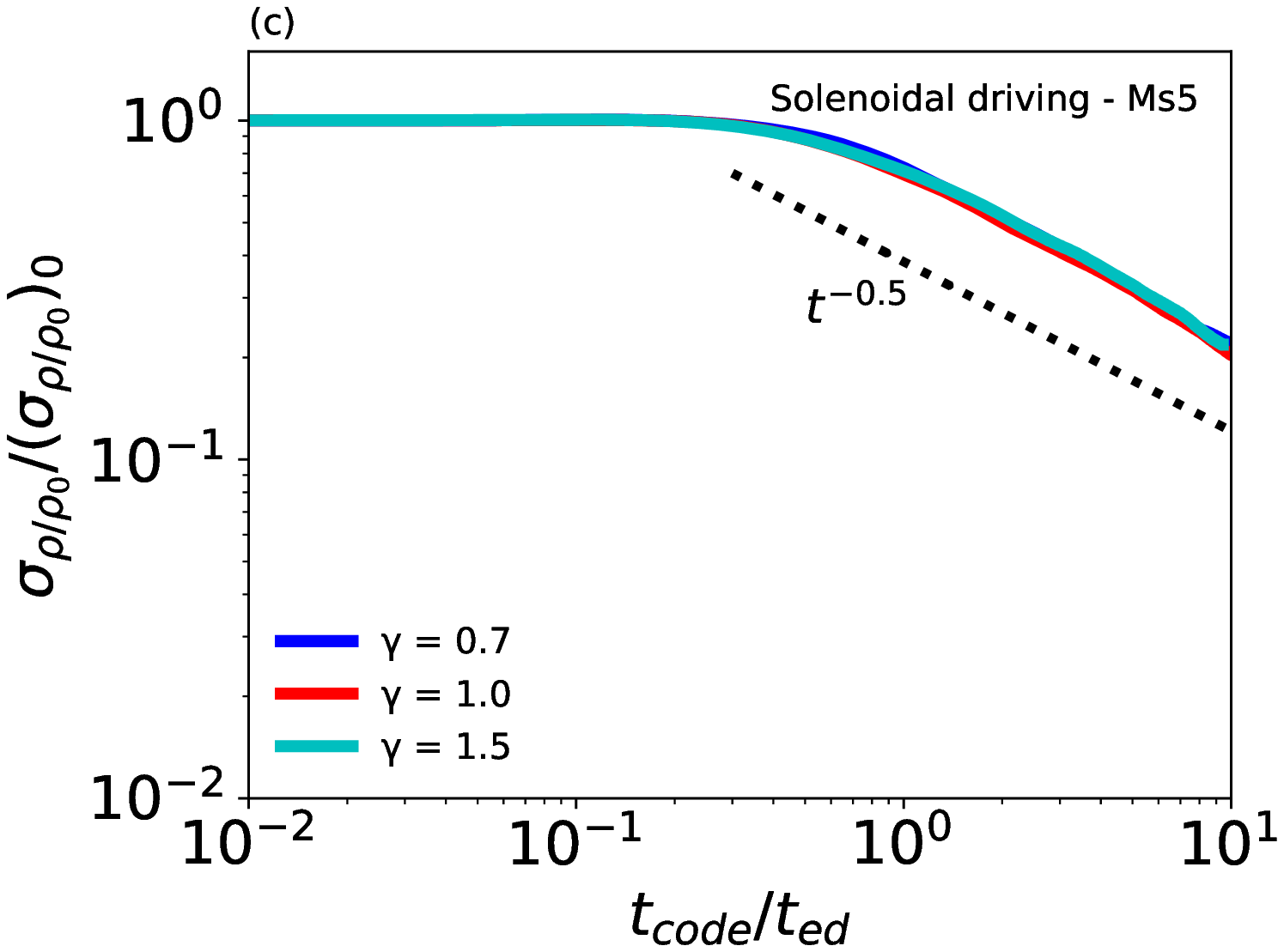}
\caption{The similar as Figure \ref{fig:fig1} but for decay of the standard deviation of the density fluctuation $\sigma_{\rho/\rho_0}$. We normalize $\sigma_{\rho/\rho_0}$ by $(\sigma_{\rho/\rho_0})_0$ which is the value at $t_{code}/t_{ed}$ = 0. \label{fig:fig2}}
\end{figure*}


\section{Motivation and Numerical Method}
\label{sec:sec2}

\subsection{Motivation\label{sec:sec2.1}}

  As we described earlier, decay of solenoidally driven isothermal turbulence follows E $\propto$ $t^{-\alpha}$ with $\alpha$ $\approx$ 1.  The type of driving or the polytropic $\gamma$ may affect this scaling relation. 
    
  First, if we use compressive driving, it yields more compressions at the same Mach number. Therefore, while decaying, compressed regions could generate additional kinetic energy via expansion, which could affect the rate of turbulence decay.
  
Second, regarding the effects of polytropic $\gamma$ on decaying turbulence, only limited parameter study is available. \citet{Mac98} found that supersonic turbulence with $\gamma$ = 1.4 decays with $\alpha$  $\sim$  1.2. For isothermal cases (i.e., $\gamma$ = 1), they found that $\alpha$ is nearly unity. This suggests that the scaling exponent $\alpha$ in E $\propto$ $t^{-\alpha}$ only weakly depends on polytropic $\gamma$ as assumed by \citet{DF17}. However, \citet{Mac98} used random initial velocity perturbation, which follows a power-law, and a constant initial density. Therefore, it is necessary to test the decay law using initial velocity and density data cubes from actual turbulence simulations with both soft EOS (i.e., polytropic $\gamma$ $<$ 1) and stiff EOS (i.e., polytropic $\gamma$ $>$ 1).

Third, the effects of polytropic $\gamma$ and the type of driving on density PDF of turbulence have also been addressed in several previous studies. For example, \citet{Fe10} showed that solenoidal driving and compressive driving can produce different statistics of isothermal turbulence as mentioned earlier. In addition, \citet{F15} found that density PDF of solenoidally driven turbulence with polytropic $\gamma$ = 5/3 has a clear power-law tail at low density, which is not observed in isothermal turbulence. However, earlier studies have not addressed turbulence with polytropic EOS and compressive driving. In this paper, we use both compressive driving and polytropic EOS to investigate density PDF and skewness of driven and decaying turbulence.

\subsection{Numerical Method\label{sec:sec2.2}}


\subsubsection{Numerical Code\label{sec:sec2.2.1}}

We use an Essentially Non-Oscillatory (ENO) scheme \citep[see][]{CL02} to solve the ideal hydrodynamic equations in a periodic box of size 2$\pi$:

\begin{equation}
\label{eq:eq2}
\frac{\partial{\rho}}{\partial{t}}+\nabla \cdot (\rho\mathbf{v})=0,
\end{equation}

\begin{equation}
\label{eq:eq3}
\frac{\partial{\mathbf{v}}}{\partial{t}}+\mathbf{v} \cdot \nabla \mathbf{v} +\rho^{-1} \nabla P =  \mathbf{f},
\end{equation}
where $\mathbf{f}$ is a driving force, $P$ is pressure (see Section \ref{sec:sec2.2.2}), $\rho$ is the density, and $\mathbf{v}$ is velocity. The density and velocity are set to 1 and zero at t = 0 to assume a static medium with a constant density at the beginning.

\begin{figure*}[ht!]
\centering
\includegraphics[scale=0.35]{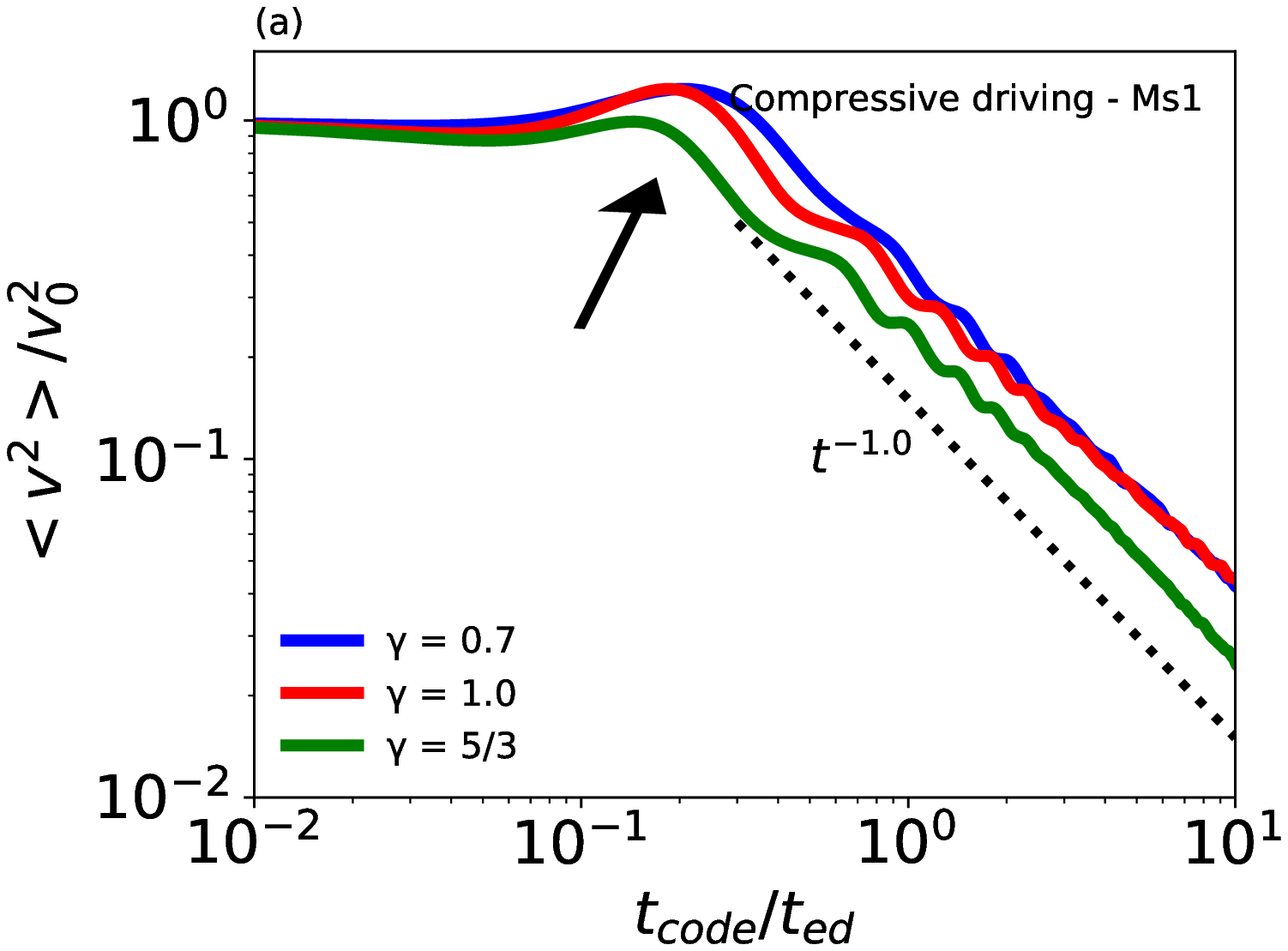}
\includegraphics[scale=0.35]{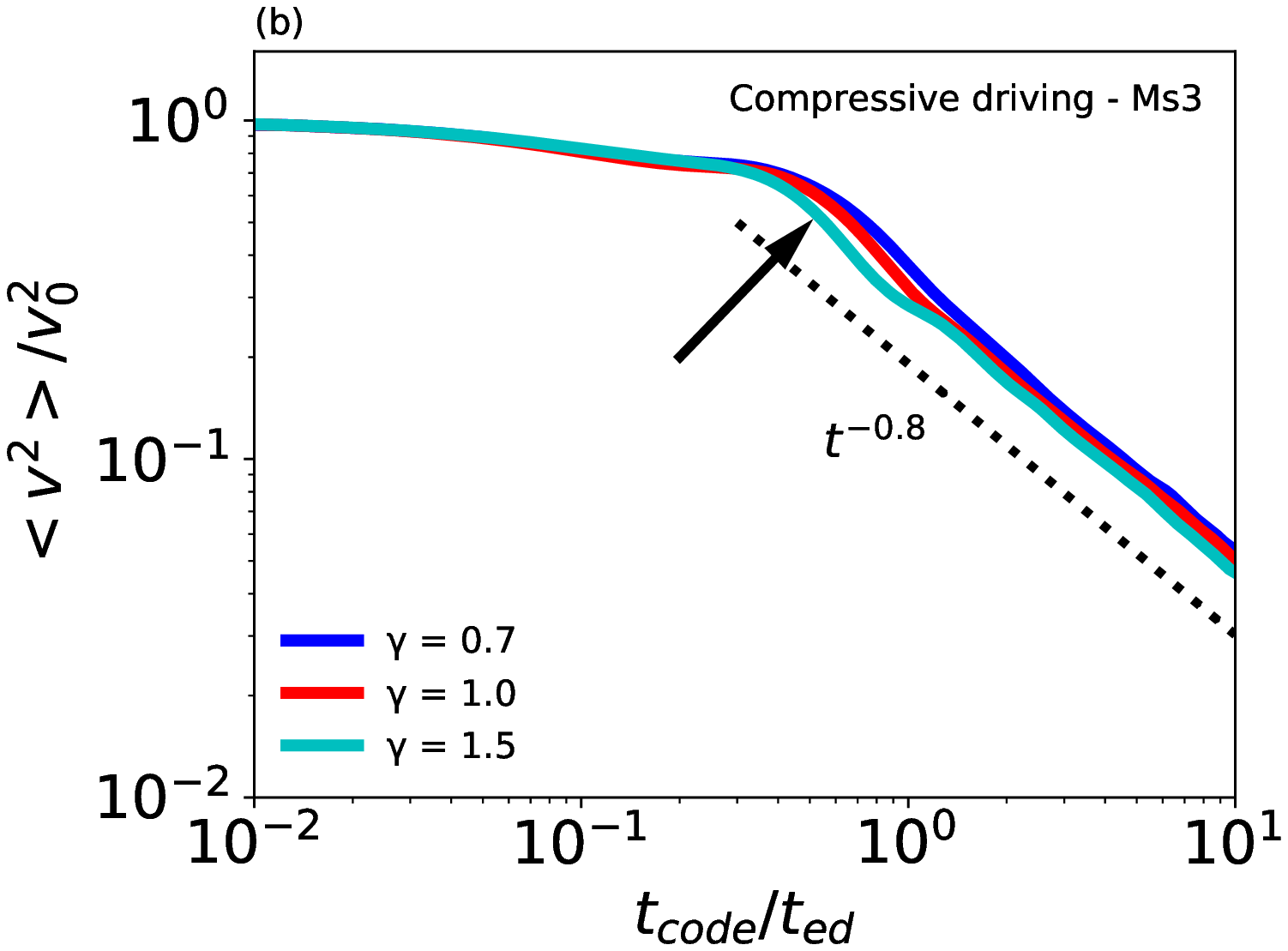}
\includegraphics[scale=0.35]{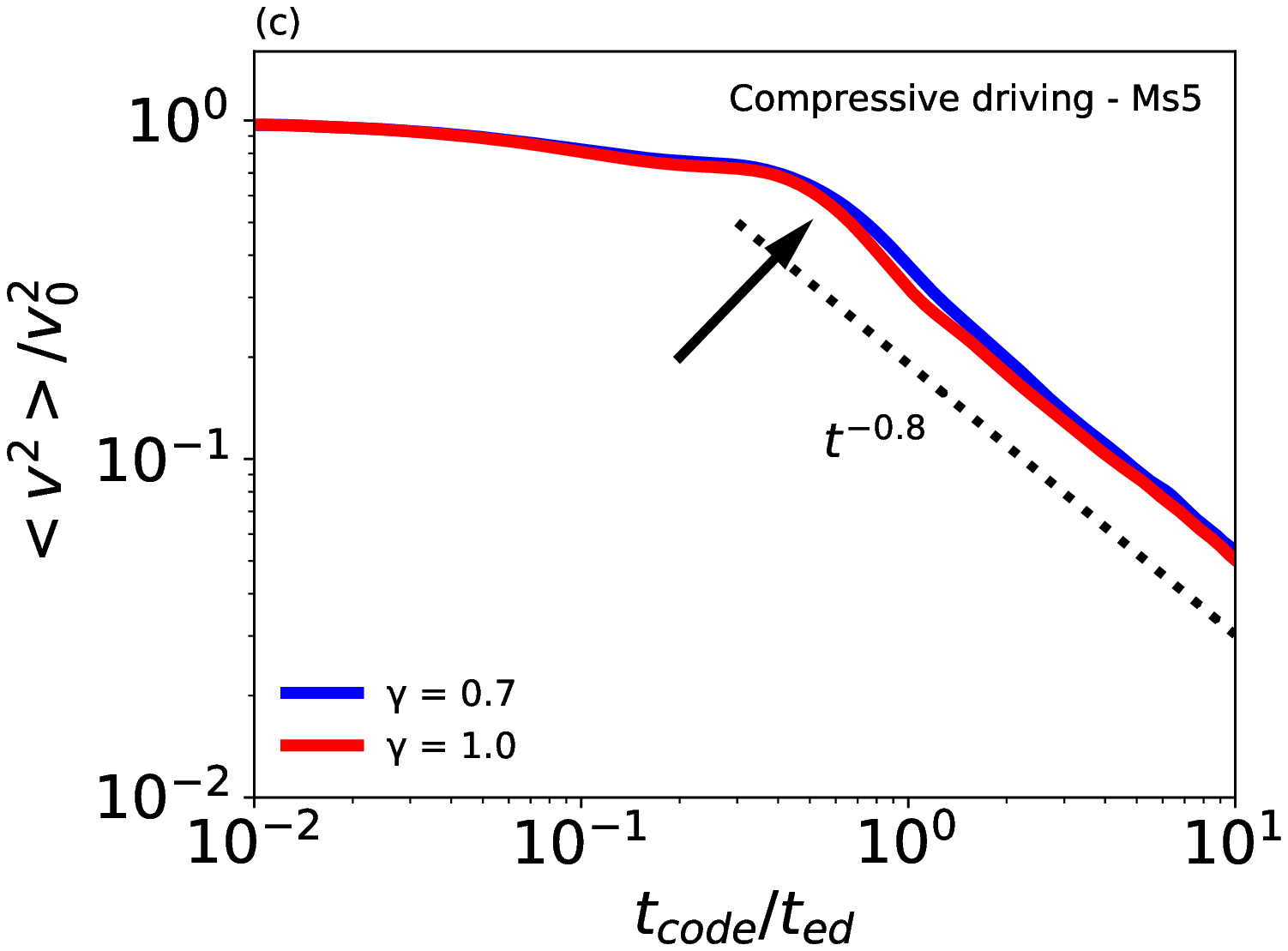}
\caption{The similar as Figure \ref{fig:fig1} but for compressively driven turbulence. Note that unlike Figure \ref{fig:fig1}, a small amount of kinetic energy densities is generated, and this is most apparent in the case of $M_{s}$ $\sim$ 1. 
\label{fig:fig3}}
\end{figure*}

\begin{figure*}[ht!]
\centering
\includegraphics[scale=0.35]{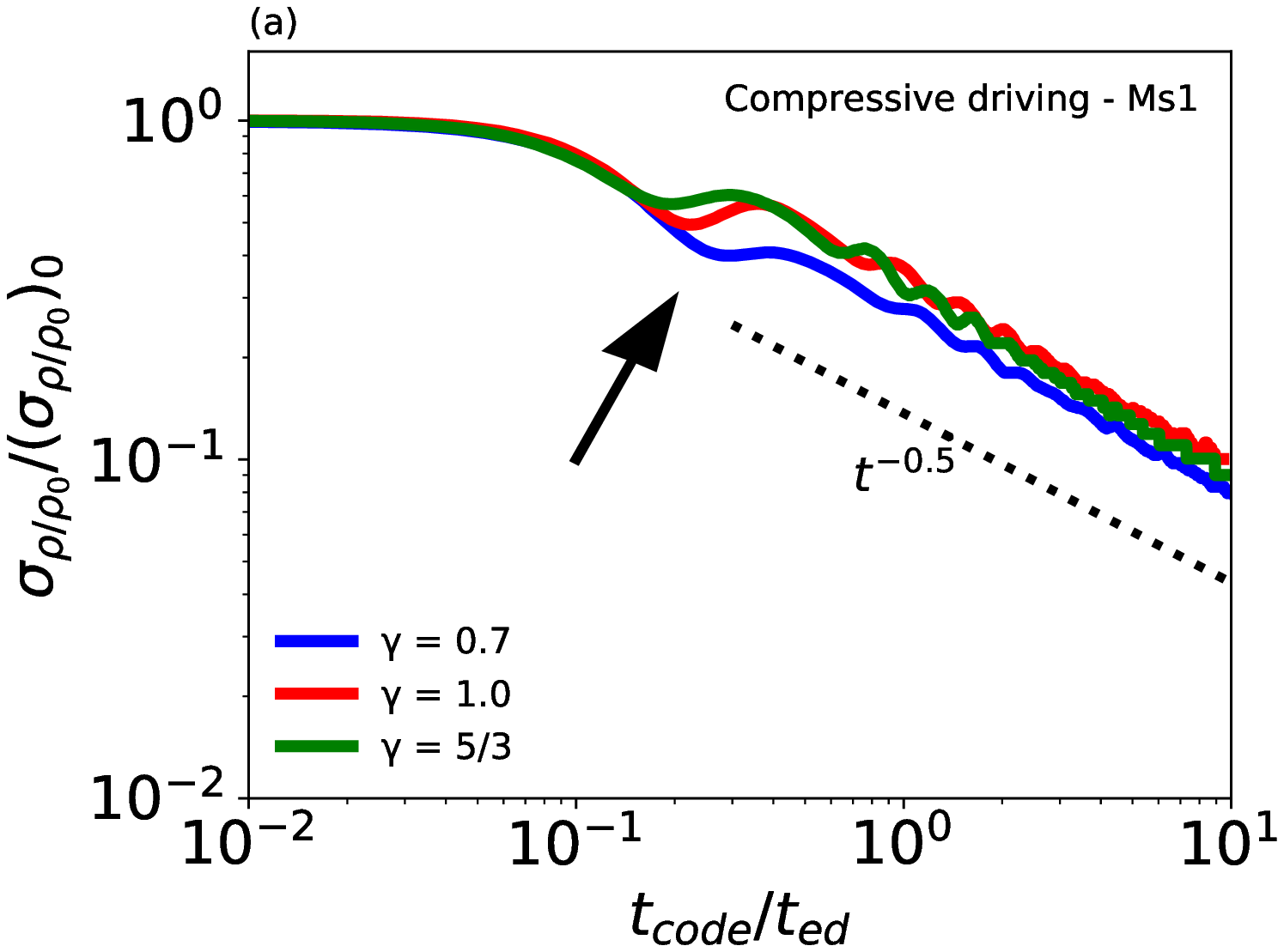}
\includegraphics[scale=0.35]{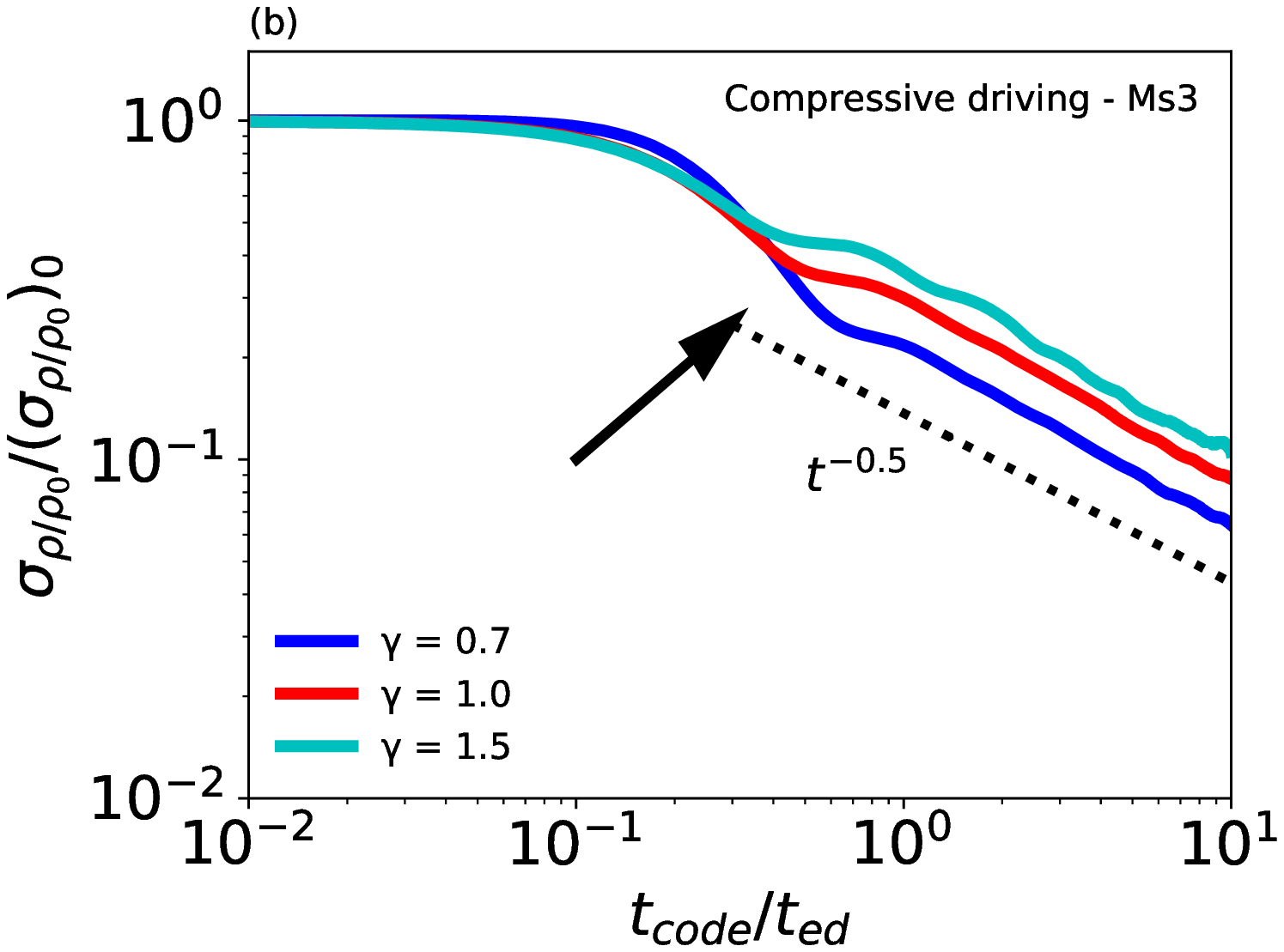}
\includegraphics[scale=0.35]{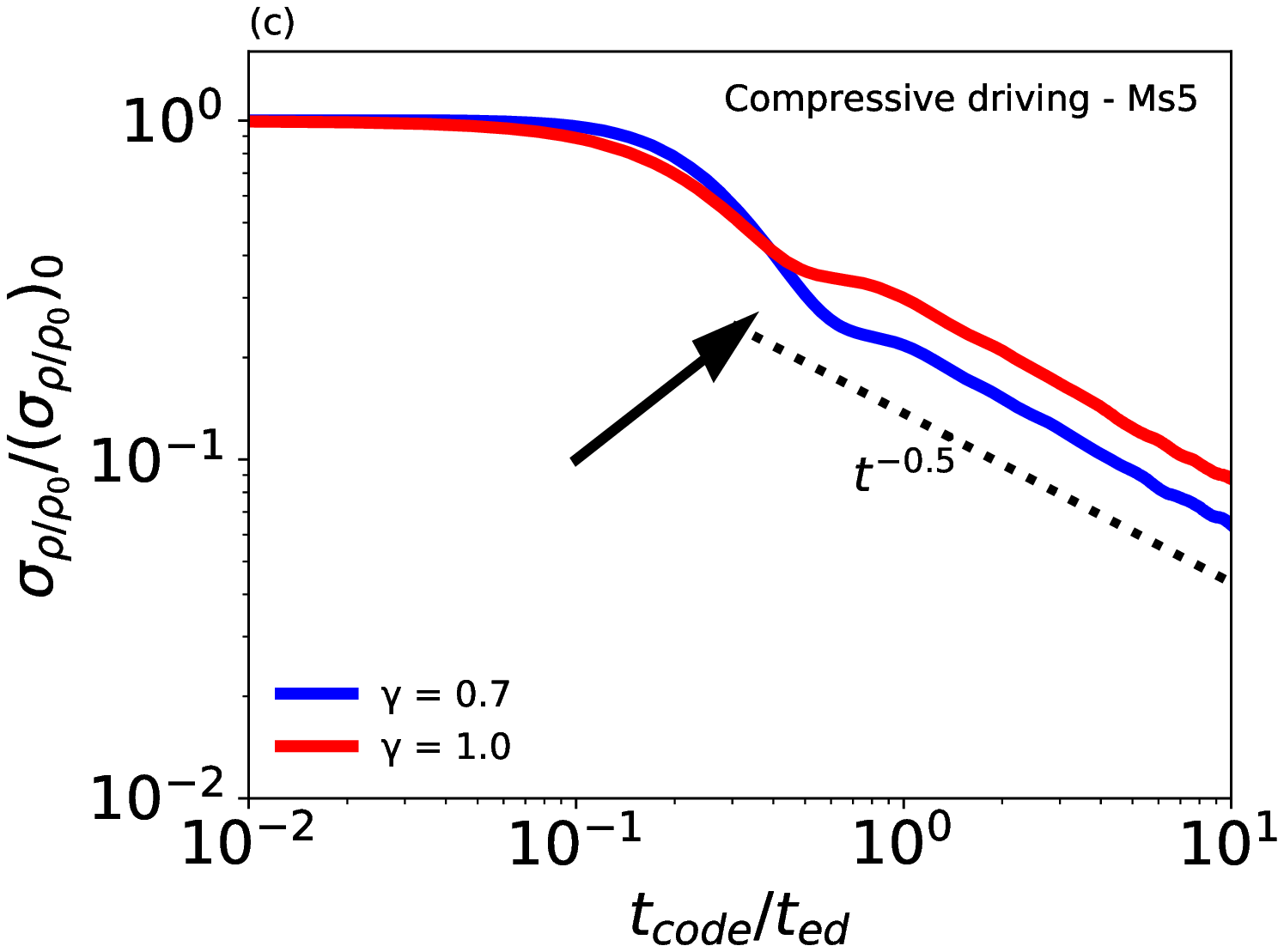}
\caption{The similar as Figure \ref{fig:fig2} but for compressively driven turbulence.
\label{fig:fig4}}
\end{figure*}


\subsubsection{Simulations\label{sec:sec2.2.2}}
We use $512^3$ grid points in our periodic computational box. The peak of energy injection occurs at $k$ $\approx$ 6 or 8, where $k$ is the wavenumber. We drive turbulence in Fourier space and use solenoidal ($\nabla \ \cdot \ \mathbf{f} = 0$) and compressive ($\nabla \ \times \ \mathbf{f} = 0$) driving. In both drivings, the driving vectors continuously change with a correlation time comparable to the large-eddy turnover time. We also adopt polytropic EOS: 
\begin{equation}\label{eq:eq4}
P = P_{0}\bigg(\frac{\rho}{\rho_0}\bigg)^{\gamma} = \bigg(\frac{c_{s0}^2\rho_0}{\gamma}\bigg)\bigg(\frac{\rho}{\rho_0}\bigg)^{\gamma}
\end{equation}
where $P$ is the normalized pressure, and $P_0$,  $c_{s0}$, and $\rho_0$ are the initial pressure, sound speed, and density, respectively. The sonic Mach number $M_s$ is defined by
\begin{equation}\label{eq:eq5}
M_{s} \equiv \frac{v_{rms}}{c_{s0}},  
\end{equation}
where $v_{rms}$ is the rms velocity. We vary the polytropic $\gamma$ and the sonic Mach number $M_{s}$ to consider both soft and stiff EOS in transonic and supersonic regimes. 

Table \ref{tab:tab1} lists our simulation models. We use the notation XMSY-$\gamma$Z, where X = S or C refers to solenoidal or compressive driving; Y = 1, 3, or 5 refers to the sonic Mach number $M_s$; Z = 0.7, 1.0, 1.5, or 5/3 refers to the value of polytropic $\gamma$. We keep driving turbulence until the system reaches saturation stage, after which the driving is turned off to let turbulence freely decay. In decaying simulations, time is normalized by $t = t_{code}/t_{ed}$. Here, $t_{code}$ is the time in code unit, and $t_{ed} = (L/k_{ej})/v_{0}$ is large-eddy turnover time, where $L$ = 2$\pi$ is the size of the simulation box, $k_{ej}$ = 6 or 8 is the driving wavenumber at which the energy injection rate peaks, and $v_{0}$ is the velocity at the moment turbulence starts decaying.

\section{Results\label{sec:sec3}}

\subsection{Decay of Hydrodynamic Turbulence with Polytropic EOS\label{sec:sec3.1}}

\subsubsection{Decay of Turbulence Driven by Solenoidal Driving\label{sec:sec3.1.1}}
In this subsection, we consider decaying polytropic turbulence initially driven by solenoidal driving.
Figures \ref{fig:fig1} and \ref{fig:fig2} show decay of kinetic energy density $<v^2>$ and standard deviation of density fluctuation $\sigma_{\rho/\rho_0}$, respectively. From left to right panel, the sonic Mach number $M_s$ is $\sim$ 1, $\sim$ 3, and $\sim$ 5, respectively. Blue, red, cyan, green curves in each panel correspond to polytropic $\gamma$ = 0.7, 1.0, 1.5, and 5/3, respectively.

First of all, we can clearly see that decay of kinetic energy density follows a power-law form of $<v^2>$ $\propto$ $t^{-\alpha}$, and $\alpha$ is almost same at the same $M_s$ regardless of the value of polytropic $\gamma$. The energy decay is steeper in the case of $M_s$ $\sim$ 1 ($\alpha$ $\sim$ 1.2) than supersonic cases ($\alpha$ $\sim$ 1.0). Second, similar to the case of $<v^2>$, the decay of $\sigma_{\rho/\rho_0}$ is hardly affected by $\gamma$. In addition, the power-law exponent for $\sigma_{\rho/\rho_0}$ is nearly half of that for $<v^2>$ at the same $M_s$. For the cases of $\gamma$ = 1 (i.e., isothermal cases), this result is consistent with the fact that standard deviation of density fluctuation is approximately linear with the sonic Mach number (e.g., \citealt{PNJ97, PV98}), which implies $\sigma_{\rho/\rho_0}$ $\propto$ $M_s$ $\propto$ $<v^2>^{1/2}$ $\propto$ $t^{-\alpha/2}$. Our results imply that a similar argument holds true for $\gamma$ $\neq$ 1.

\subsubsection{Decay of Turbulence Driven by Compressive Driving\label{sec:sec3.1.2}}

Now, let us deal with decay of polytropic turbulence initially driven by compressive driving. Figures \ref{fig:fig3} and \ref{fig:fig4} show decay of $<v^2>$ and $\sigma_{\rho/\rho_0}$, respectively. As in the case of solenoidal turbulence\footnote{We mean solenoidal turbulence by turbulence initially driven by solenoidal driving.}, we use different values of $M_s$ (from left to right panel) and polytropic $\gamma$ (curves with different colors).

Similar to the result from solenoidal turbulence, both $<v^2>$ and $\sigma_{\rho/\rho_0}$ in compressively driven turbulence also exhibit power-law decay, and polytropic $\gamma$ hardly affects the decay rate. According to Figure \ref{fig:fig3}, the power-law exponent $\alpha$ is $\sim$ 1.0 for $M_s$ $\sim$ 1, and $\sim$ 0.8 for $M_s$ $>$ 1, which means that decay of compressively driven turbulence is slower than that of solenoidal turbulence at the same $M_s$. As can be seen from Figure \ref{fig:fig4}, the power-law exponent $\alpha$ for the decay of $\sigma_{\rho/\rho_0}$ is nearly half of that for $<v^2>$, which is consistent with the result from the previous section.

Note that, unlike the case of solenoidal turbulence, polytropic $\gamma$ slightly affects the decay of compressively driven turbulence. First, kinetic energy density in compressively driven turbulence shows bump-like features (indicated by the black arrow in each panel in Figure \ref{fig:fig3}). This slight increase of kinetic energy density is most pronounced in the case of $M_s$ $\sim$ 1. Second, dip-like features (indicated by the black arrow in each panel in Figure \ref{fig:fig4}) are clearly shown in the evolution of $\sigma_{\rho/\rho_0}$ at the nearly same time when the bump-like features in $<v^2>$ occur. Third, we can see from Figure \ref{fig:fig4} that decay of $\sigma_{\rho/\rho_0}$ for $\gamma$ = 0.7 is faster than that obtained for $\gamma$ $>$ 0.7 regardless of $M_s$.

\begin{figure*}[]
\centering
\includegraphics[scale=0.35]{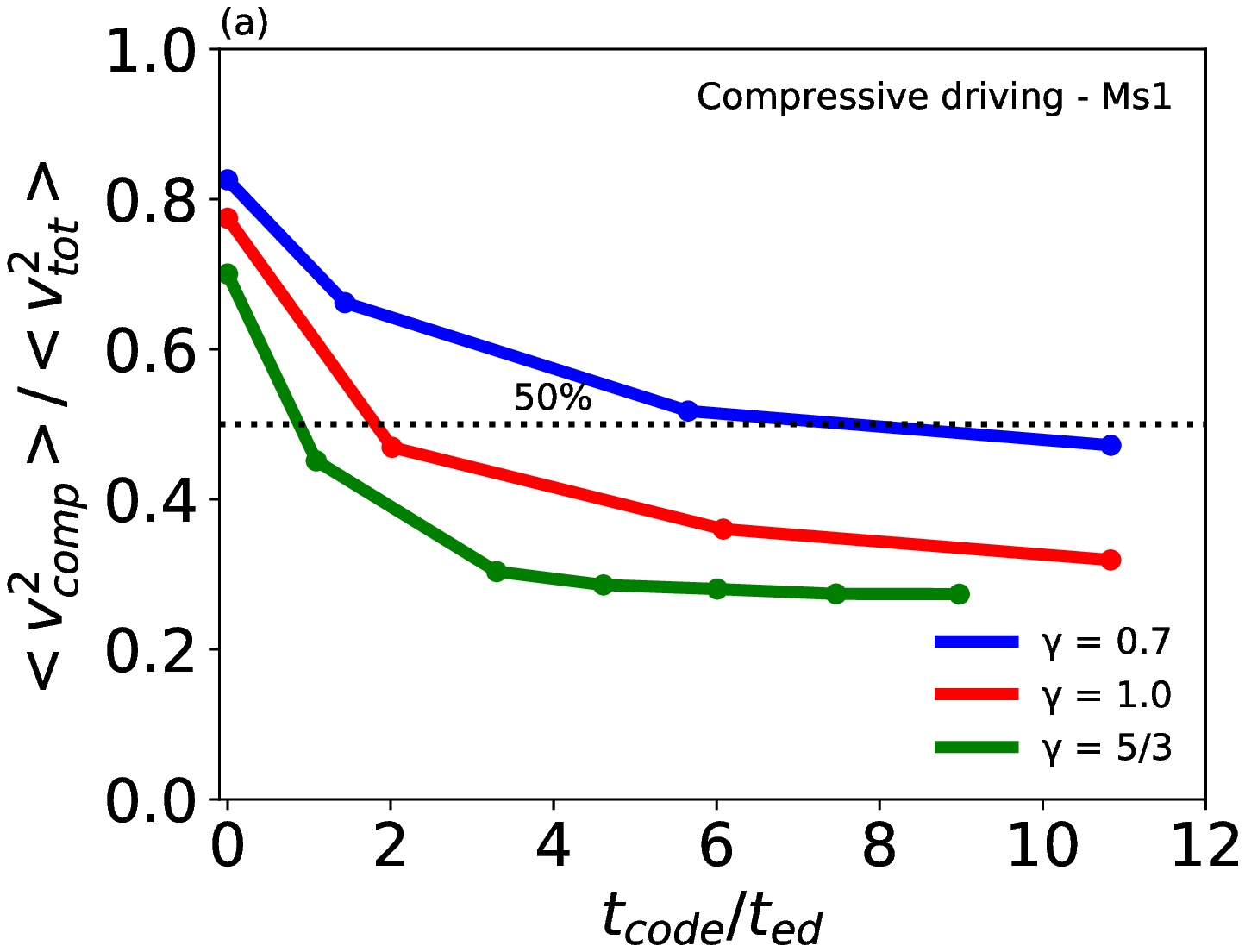}
\includegraphics[scale=0.35]{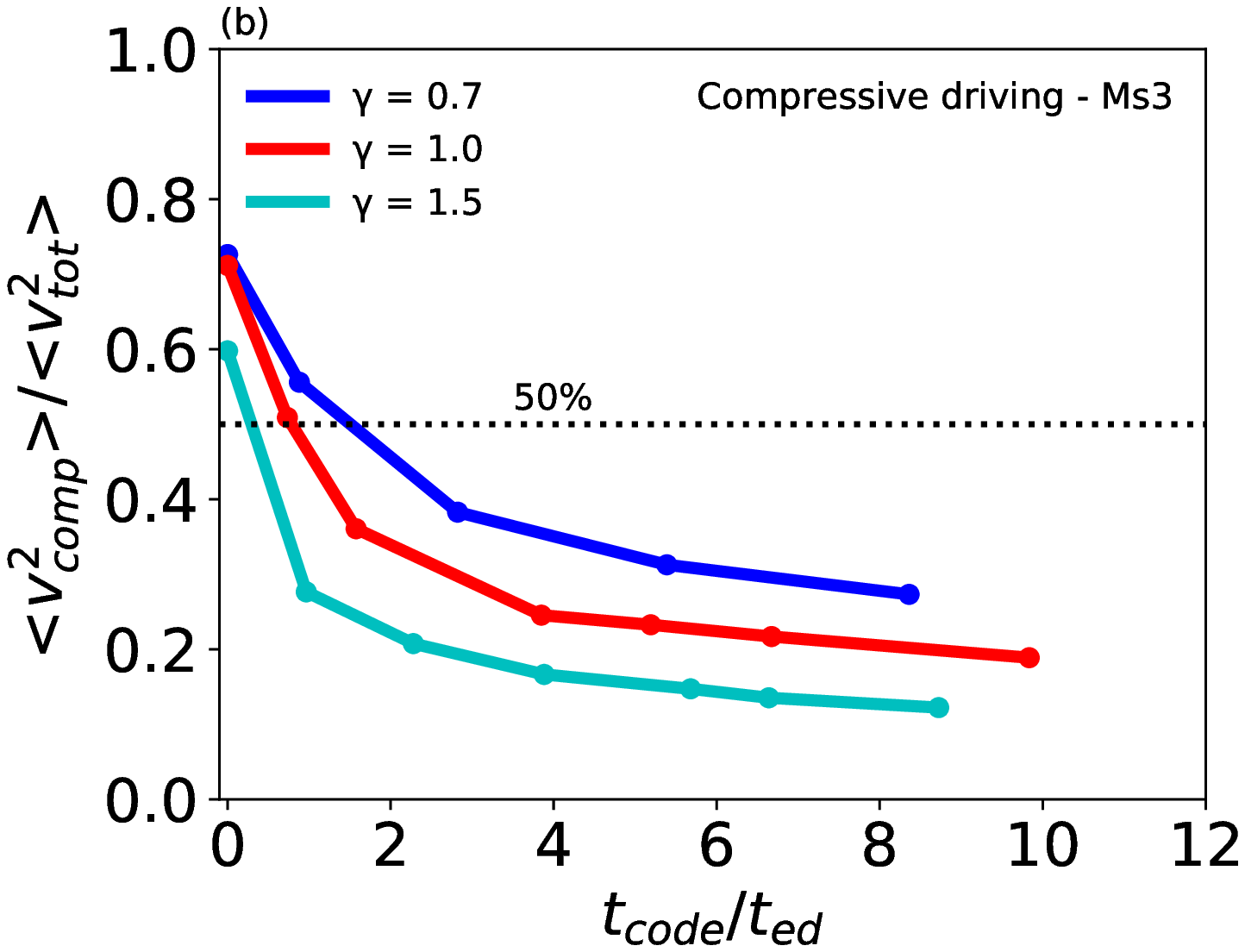} 
\includegraphics[scale=0.35]{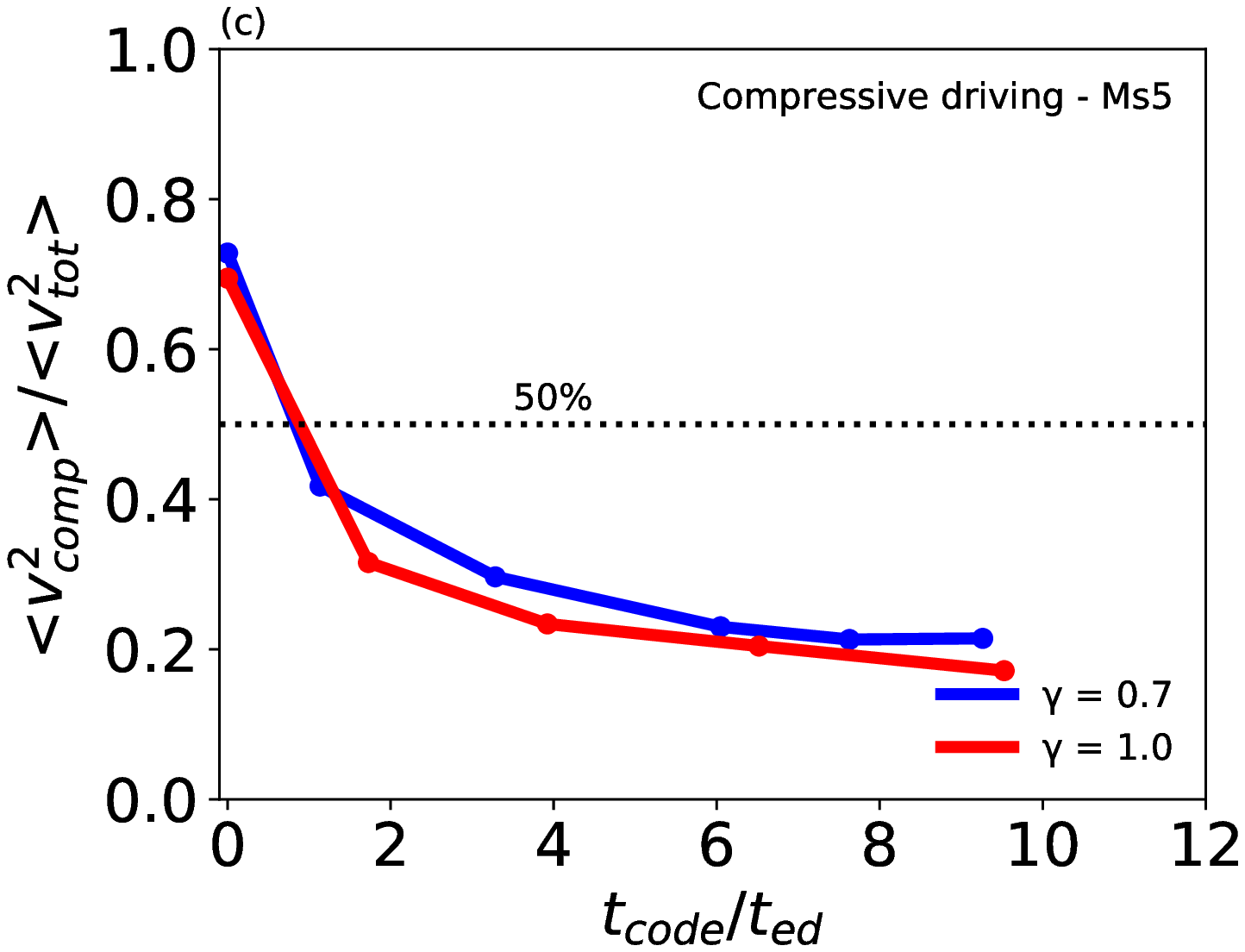} \\
\includegraphics[scale=0.35]{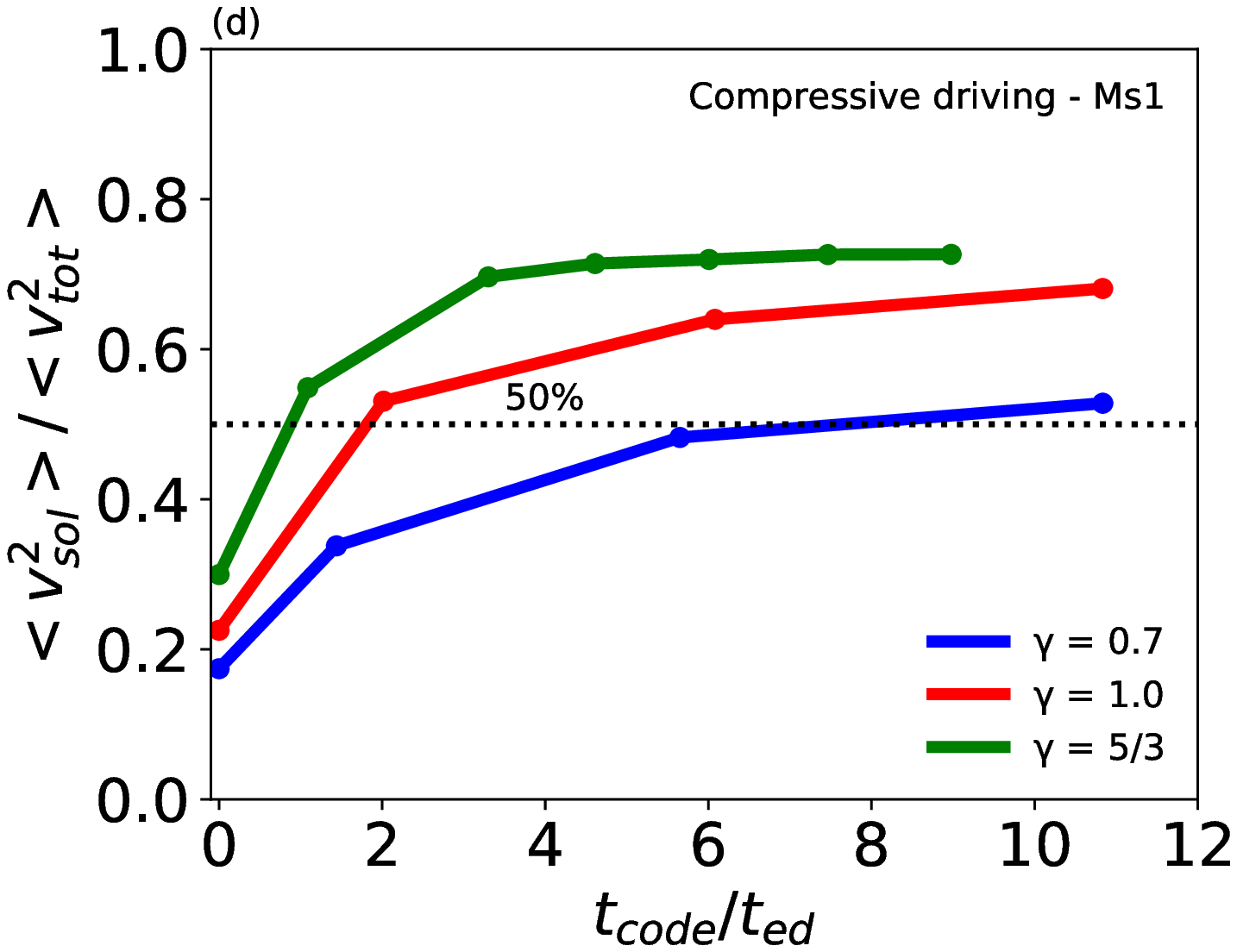} 
\includegraphics[scale=0.35]{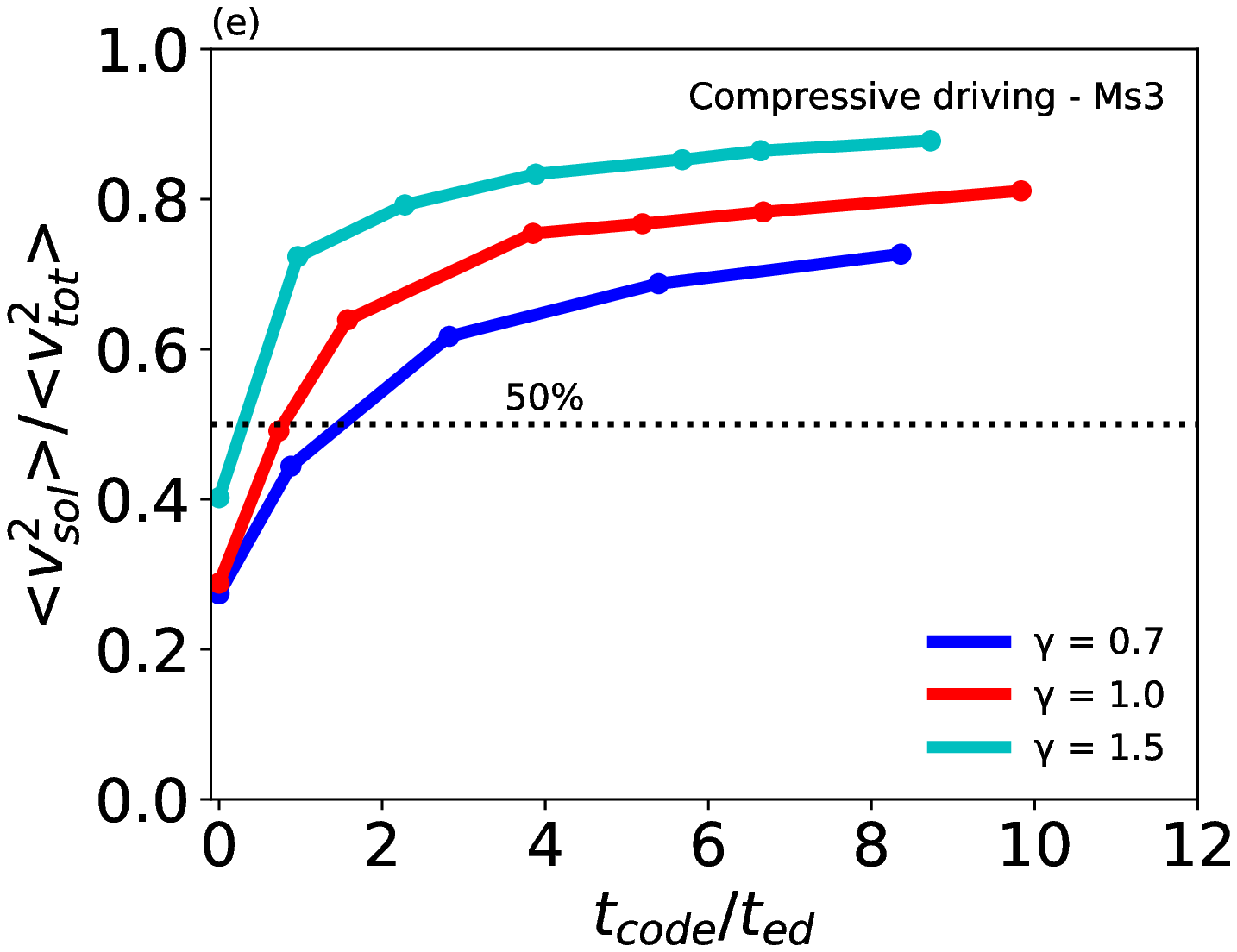}
\includegraphics[scale=0.35]{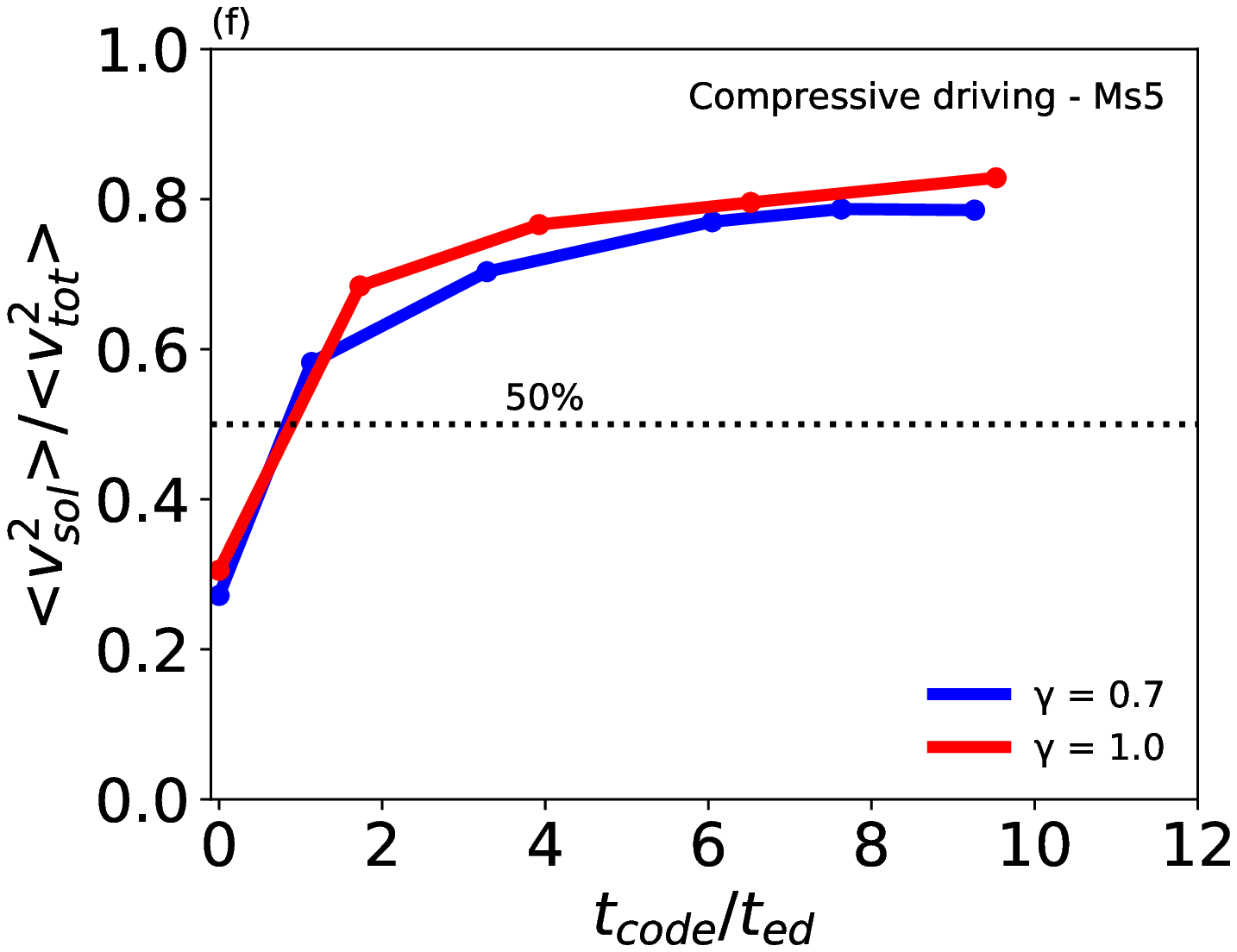}
\caption{Decay of ratio of both compressive (upper panels) and solenoidal (bottom panels) energy density in compressively driven turbulence. Left panels: $M_s \ \sim \ 1$. Middle panels: $M_s \ \sim \ 3$. Right panels: $M_s \ \sim \ 5$. Blue, red, cyan, and green curves denote polytropic $\gamma$ = 0.7, 1.0, 1.5, and 5/3, respectively. 
\label{fig:fig5}}
\end{figure*}

\begin{figure*}[t!]
\centering
\includegraphics[scale=0.35]{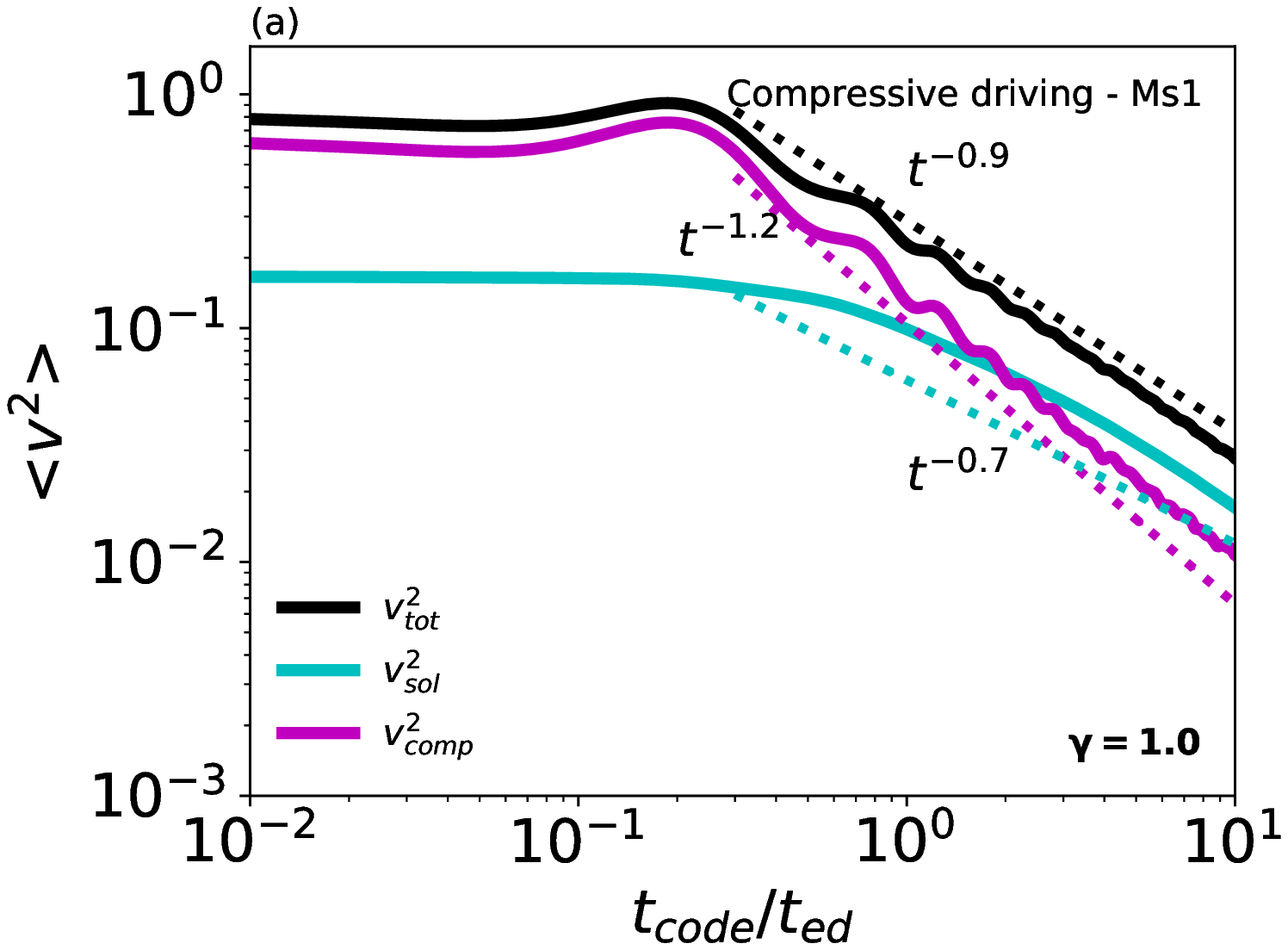}
\includegraphics[scale=0.35]{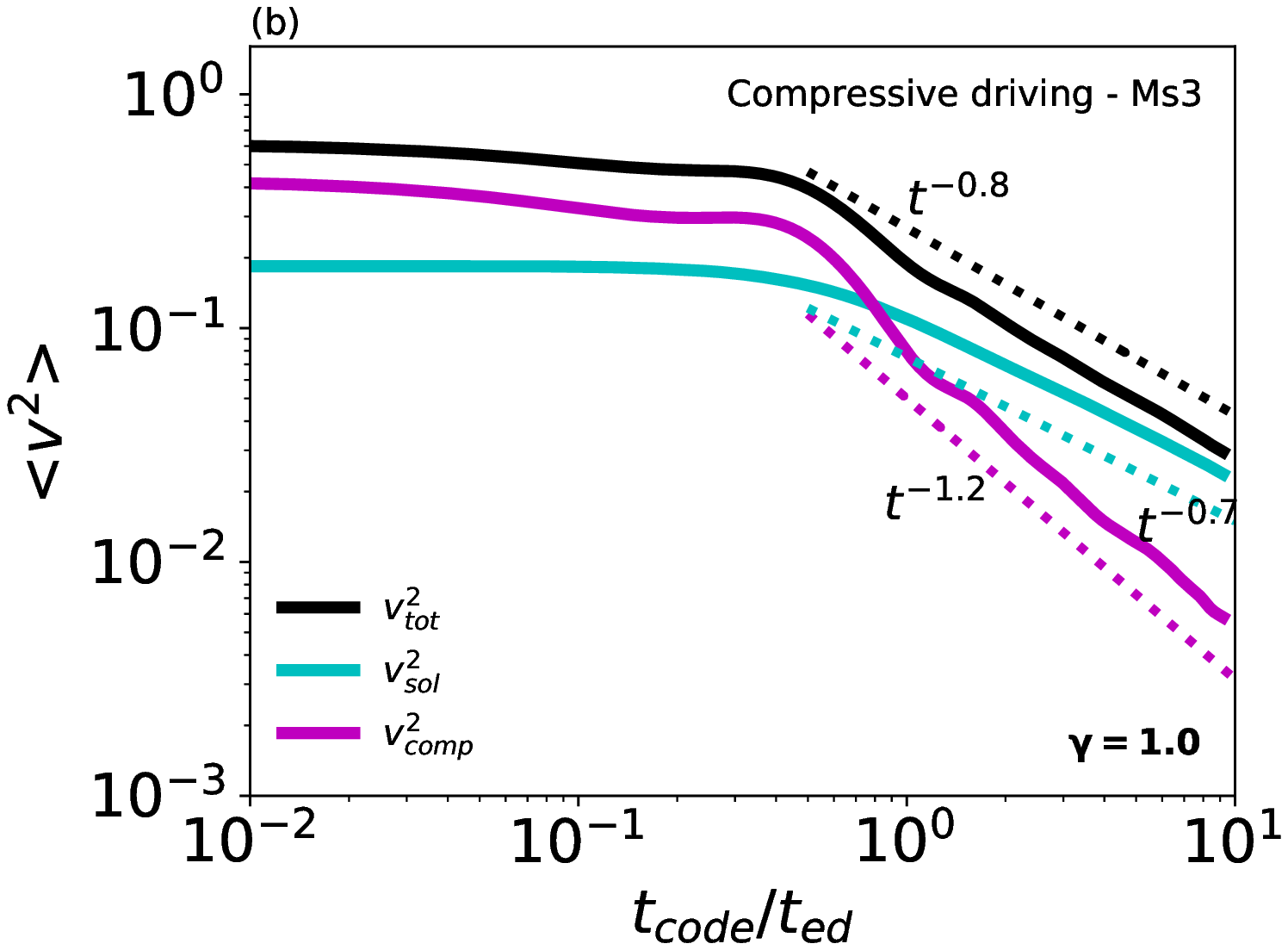}
\includegraphics[scale=0.35]{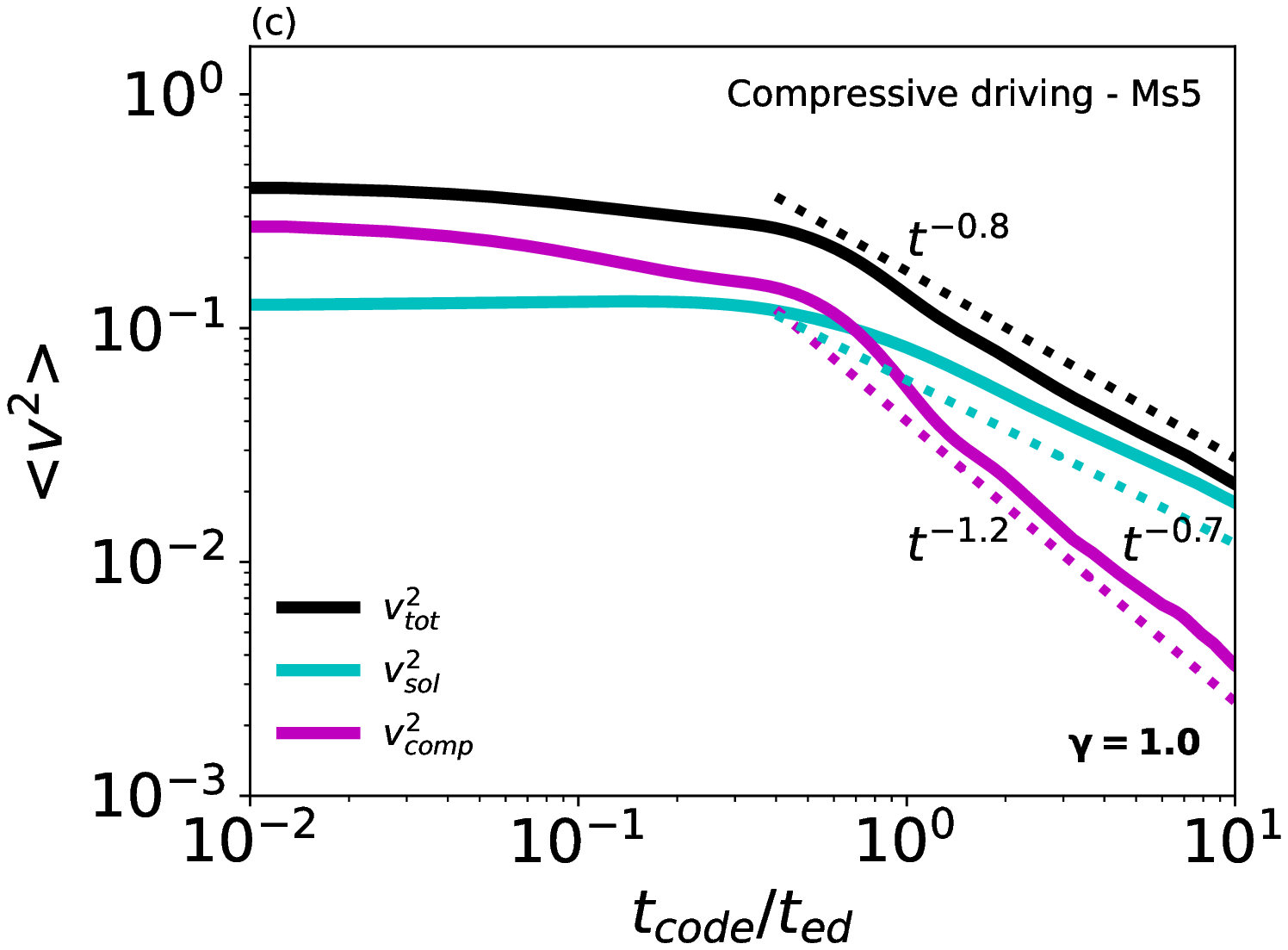}
\caption{Decay of both solenoidal and compressive energy density in compressively driven turbulence. Only isothermal turbulence (i.e., polytropic $\gamma$ = 1.0) is presented in this figure. Left panel: $M_s \ \sim \ 1$. Middle panel: $M_s \ \sim \ 3$. Right panel: $M_s \ \sim \ 5$. Black, cyan, and magenta curves represent total, solenoidal, and compressive energy density, respectively. The dotted lines with different colors are reference lines for different power-law forms. Note that compressive kinetic energy density decays much faster than solenoidal one.
\label{fig:fig6}}
\end{figure*}

\subsection{Decay of Solenoidal and Compressive Velocity Components in Compressively Driven Turbulence\label{sec:sec3.2}}

In next two subsections, we only consider compressively driven turbulence. In this subsection, we first deal with how differently solenoidal and compressive modes decay. In order to address the issue, we decompose 3D velocity fields of the compressively driven turbulence into solenoidal and compressive components. Figure \ref{fig:fig5} illustrates the decay of compressive ratio $<v_{comp}^2>/<v_{tot}^2>$ (upper panels) and solenoidal ratio $<v_{sol}^2>/<v_{tot}^2>$ (bottom panels). Here, $v_{tot}^2$ = $v_{sol}^2$ + $v_{comp}^2$, and $v_{sol}^2$ and $v_{comp}^2$ are solenoidal and compressive kinetic energy density, respectively. Blue, red, cyan, and green curves indicate polytropic $\gamma$ = 0.7, 1.0, 1.5, and 5/3, respectively.

First, Figure \ref{fig:fig5} shows that the compressive ratio decreases as turbulence decays. When $M_s$ $\sim$ 1 or $\sim$ 3, the smaller the polytropic $\gamma$ is, the larger the compressive ratio is. However, when $M_s$ $\sim$ 5, we do not see strong dependence of the compressive ratio on $\gamma$. Second, and more importantly, the solenoidal ratio increases as turbulence decays and eventually becomes higher than the compressive ratio irrespective of $\gamma$ and $M_s$, which means that compressive energy density decays faster. Figure \ref{fig:fig6} clearly shows this in the case of isothermal turbulence initially driven by compressive driving, in which we plot the decay of solenoidal and compressive energy density separately. We can clearly see from the reference lines (see dotted lines with different colors in each panel) that compressive energy density (magenta curves) decays more quickly than solenoidal energy density (cyan curves), with this resulting in higher solenoidal energy density at the late stages of decay irrespective of $M_s$.

\begin{figure*}[ht!]
\centering
\includegraphics[scale=0.75]{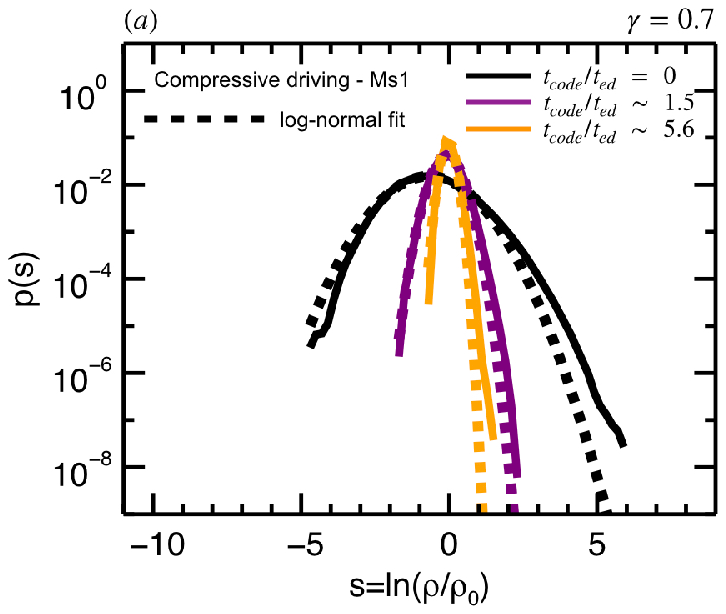}
\includegraphics[scale=0.75]{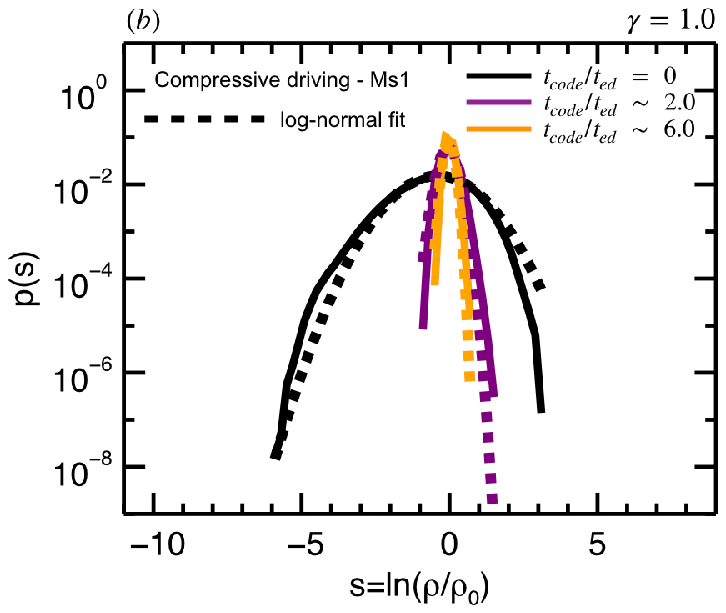} 
\includegraphics[scale=0.75]{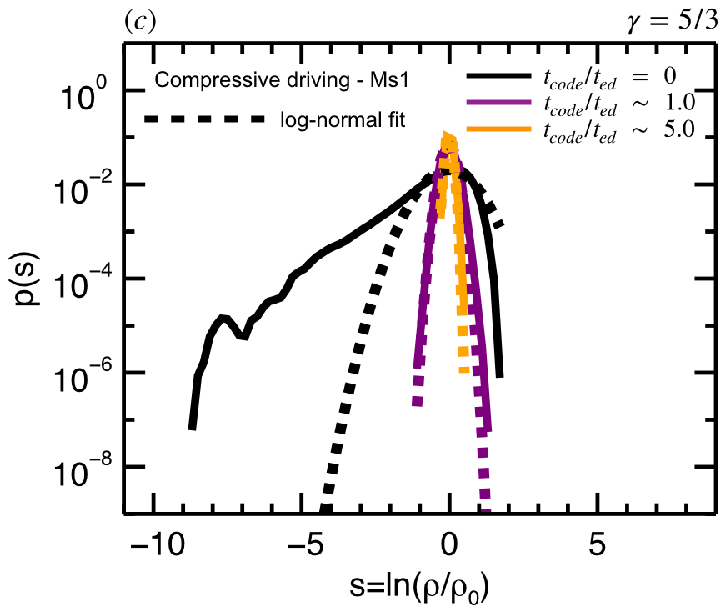} 
\caption{Density PDF of the logarithmic density $s=ln(\rho/\rho_0)$ of decaying polytropic turbulence initially driven by compressive driving with $M_{s}$ $\sim$ 1. Left panels: polytropic $\gamma$ = 0.7. Middle panels: $\gamma$ = 1.0. Right panels: $\gamma$ = 5/3. The y-axis is logarithmic scale. Black, purple, and orange curves in each panel represent different times along the decay. The dotted line in each panel is log-normal fitting line.
\label{fig:fig7}}
\end{figure*}

\begin{figure*}[h!]
\centering
\includegraphics[scale=0.75]{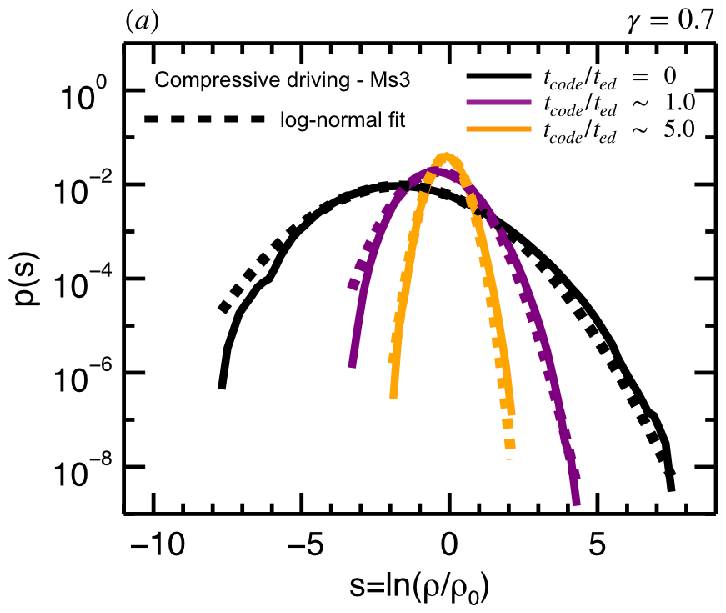}
\includegraphics[scale=0.75]{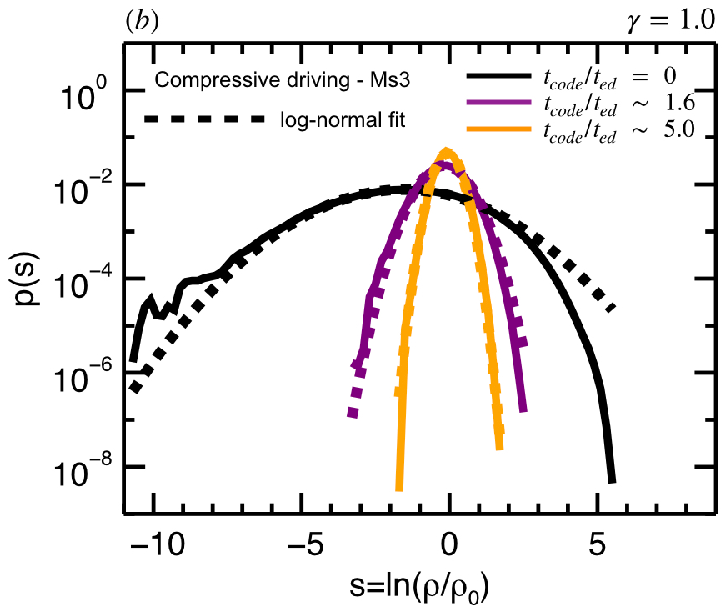} 
\includegraphics[scale=0.75]{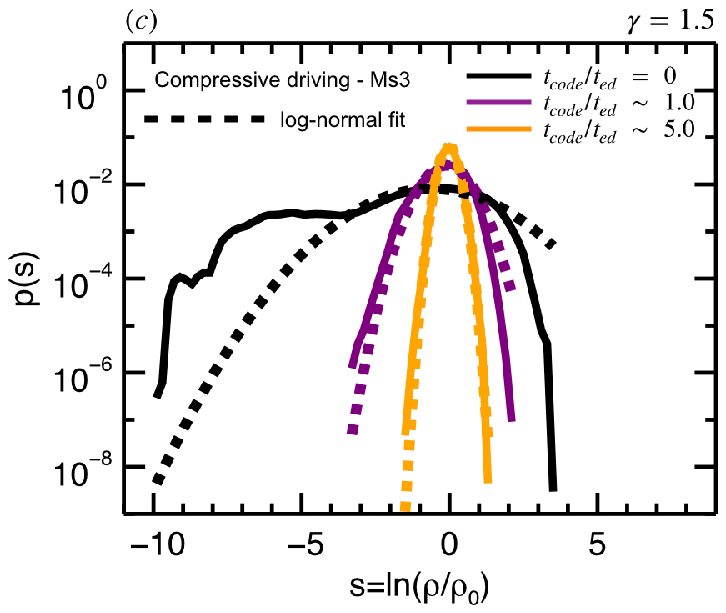}  
\caption{The similar as Figure \ref{fig:fig7} but for $M_{s}$ $\sim$ 3 cases. Left panels: polytropic $\gamma$ = 0.7. Middle panels: $\gamma$ = 1.0. Right panels: $\gamma$ = 1.5. 
\label{fig:fig8}}
\end{figure*}

\begin{figure*}[ht!]
\centering
\includegraphics[scale=0.5]{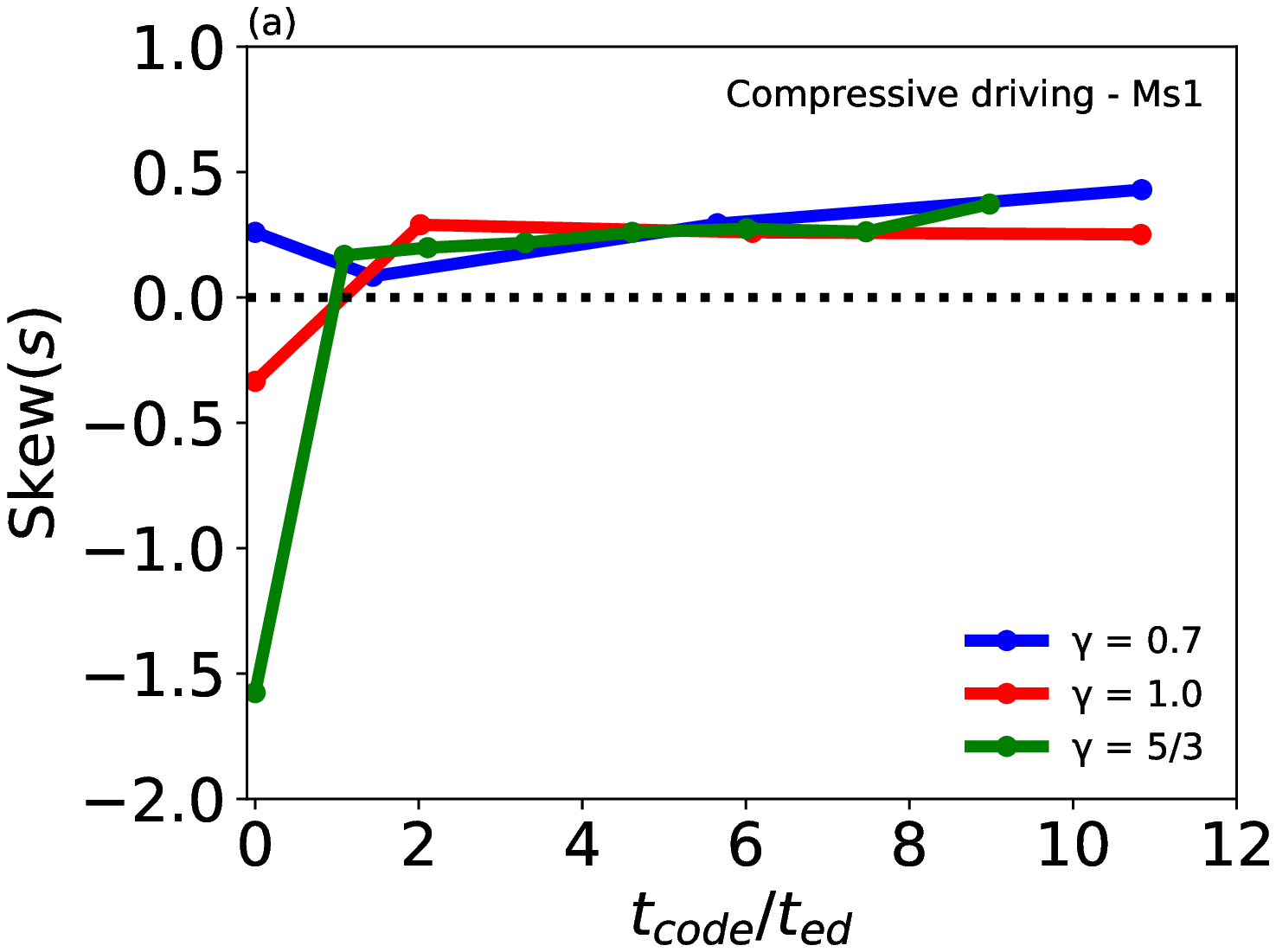}
\includegraphics[scale=0.5]{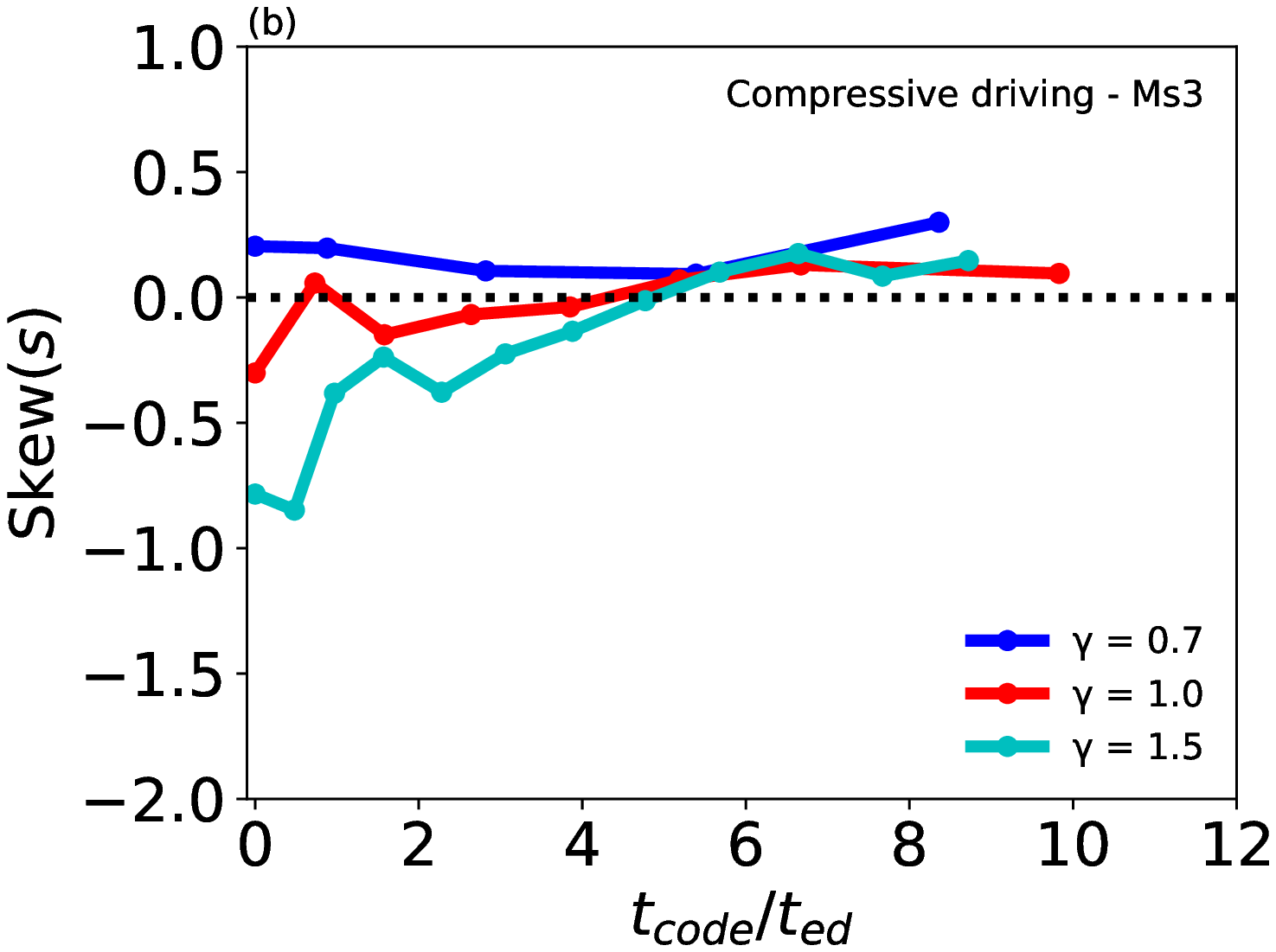}
\caption{Time evolution of skewness of the density PDFs shown in Figure \ref{fig:fig7} (left panel) and Figure \ref{fig:fig8} (right panel). Blue, red, cyan, green curves correspond to polytropic $\gamma$ = 0.7, 1.0, 1.5, and 5/3, respectively.
\label{fig:fig9}}
\end{figure*}

\subsection{Density PDF and Skewness\label{sec:sec3.3}}
In this subsection, we investigate density PDF and its skewness of compressively driven turbulence with polytropic EOS in driven and decay regime. We define skewness of the density PDF as follow:
\begin{equation}
\label{eq:eq6}
\textnormal{Skew(s)} = \frac{1}{N}\sum_{i=1}^{N}{\big(\frac{s_i-<s>}{\sigma_s}\big)^3} 
\end{equation}

where N is the total number of data points, $s$ $\equiv$ $ln(\rho/\rho_0)$ is the natural logarithm of the density fluctuation, and $<\cdots>$ denotes the spatial average value. Skewness measures asymmetry of a probability distribution. When the distribution is left (right)-skewed, skewness has a negative (positive) value. 


Figures \ref{fig:fig7} and \ref{fig:fig8} show the density PDF of polytropic turbulence with $M_s$ $\sim$ 1 and $\sim$ 3, respectively. Each solid curve with different colors in each panel corresponds to the density PDF at different times along the decay. We carry out log-normal fitting for the PDFs using the equation:
\begin{equation}
\label{eq:eq8}
p_s(s) = \frac{1}{\sqrt{2\pi \sigma_s}}\textnormal{exp}\Big[-\frac{(s-<s>)^2}{2\sigma_s^2}\Big].	
\end{equation}
The fitting line is indicated as the dotted lines in each panel.

First, let us consider driven turbulence as indicated by black solid line in Figures \ref{fig:fig7} and \ref{fig:fig8}. As we can see, compressively driven turbulence yields density PDFs which are not perfectly log-normal even in the case for $\gamma$ = 1. The density PDF for $\gamma$ = 0.7 is slightly right-skewed, and that for $\gamma$ = 5/3 is strongly left-skewed. The latter has a pronounced power-law tail at low density as shown in Figure \ref{fig:fig7}(c). Figure \ref{fig:fig8}(c) shows that the PDF of CMS3-$\gamma$1.5, which is for $M_s$ $\sim$ 3, deviates more strongly from the log-normal form at low density. However, it is not clear whether the low density tail follows a power-law.
 
 Note that our results are for compressively driven turbulence with polytropic EOS. Earlier studies are available for solenoidal turbulence with polytropic EOS and $M_s$ $\sim$ 10 \citep{F15}, and also for compressively driven turbulence with $\gamma$ = 1 and $M_s$ $\sim$ 5 \citep{Fe10}. Our current result is consistent with that of the latter reference in that the PDF is slightly left-skewed when $\gamma$ = 1. Our result is also consistent with that of the former reference in that the PDF is strongly left-skewed with a power-law tail when $\gamma$ = 5/3. However, the PDF of solenoidal turbulence is more or less symmetric when $\gamma$ = 0.7 or 1.0 \citep{F15}, while that of compressively driven turbulence is clearly right-skewed for transonic turbulence and slightly right-skewed in supersonic turbulence when $\gamma$ = 0.7 (see Figures \ref{fig:fig7}(a) and \ref{fig:fig8}(a)).
 
Figure \ref{fig:fig9} shows time evolution of skewness of density PDF of compressively driven turbulence. The horizontal axis of the figure denotes the elapsed decay time normalized by the large-eddy turnover time. The left and right panels of Figure \ref{fig:fig9} show skewness for $M_s$ $\sim$ 1 and $\sim$ 3, respectively. At t = 0, as we can see from the figure, the PDF for $\gamma$ $>$ 1 has negative skewness (see green curve in the left panel and cyan curve in the right panel). Skewness for other values of $\gamma$ also indicates that the density PDFs presented deviate from log-normal form at t = 0.

Next, we consider decay regime. As can be seen from purple and orange solid curves in Figures \ref{fig:fig7} and \ref{fig:fig8}, as turbulence decays, the density PDFs become narrow and get close to log-normal forms in all cases. We can confirm this trend in Figure \ref{fig:fig9}. Skewness for $M_s$ $\sim$ 1 and $\sim$ 3 cases approaches and fluctuates around zero as turbulence decays, which is consistent with the temporal change of the PDFs shown in Figures \ref{fig:fig7} and \ref{fig:fig8}.


\section{Discussion and Summary\label{sec:sec4}}
The purpose of this study is to investigate the effects of EOS (i.e., value of polytropic $\gamma$) and driving schemes (i.e., solenoidal and compressive driving) on decaying turbulence and its statistics. In this paper, it is proved that the scaling relation of the decay law ($<v^2>$ $\propto$ $t^{-\alpha}$) does not show strong dependence on $\gamma$ and the driving schemes. Throughout the whole simulations, the kinetic energy density decays with 0.8 $\lesssim$ $\alpha$ $\lesssim$ 1.2. The range is nearly same as what \citet{Mac98} found (0.85 $<$ $\alpha$ $<$ 1.2). 

For polytropic $\gamma$ $>$ 1 cases, $\alpha$ ranges from 1.0 to 1.2 in solenoidal turbulence and from 0.8 to 1.0 in compressively driven turbulence, with the largest value of $\alpha$ being obtained in the case of $M_s$ $\sim$ 1 in both driving schemes. This result confirms the assumption of \citet{DF17} that for $\gamma$ = 5/3, $\alpha$ falls into the range 1.0 $\sim$ 1.5 with slight dependence on initial Mach number.

Even if no relationship between polytropic $\gamma$ and scaling relation of the decay law ($<v^2>$ $\propto$ $t^{-\alpha}$) is found through our study, there are several noticeable characteristics in the case of compressively driven turbulence. First, the slight increase of $<v^2>$ and the associated decrease of $\sigma_{\rho/\rho_0}$ are found in Figures \ref{fig:fig3} and \ref{fig:fig4}, respectively. As we described earlier, those effects can be interpreted as the additional energy released from compressed regions via expansion. Second, in the case of $M_s$ $\sim$ 1, the effect is most significant. This is possibly due to relatively strong pressure compared to that of supersonic cases, which results in the stronger expansion. Third, when $M_s$ is same, the bump and dip like features are more prominent in the case of $\gamma$ = 0.7. This is because when compressive driving is applied, turbulent gas with $\gamma$ = 0.7 is more easily compressed. Thus, they are also easily expanding as turbulence decays, which leads to the clearer feature in the case of $\gamma$ = 0.7. Lastly, $\sigma_{\rho/\rho_0}$ decays more quickly for $\gamma$ = 0.7 than for $\gamma$ $>$ 0.7 in both transonic and supersonic turbulence driven by compressive driving. The reason is that when $\gamma$ is less than one, expansion increases internal temperature, which dissipates density structures quickly. Therefore, density decays faster for smaller $\gamma$.

More interestingly, compressive energy density decays faster than solenoidal energy density in the case of turbulence initially driven by compressive driving as shown in Section \ref{sec:sec3.2}. We can interpret this as follows. When turbulence initially driven by compressive driving decays, the energy of compressive component is dissipated through both turbulent cascade and the dissipation at shocks, and a fraction of the energy would convert into solenoidal energy. On the contrary, it would be only turbulent cascade that allows energy of solenoidal component to be dissipated. Therefore, because of less channels for energy dissipation and the contribution from compressive component, solenoidal energy density in turbulence initially driven by compressive driving decays slower. However, more detailed analysis, such as what fraction compressive kinetic energy changes into solenoidal kinetic energy, is limited and beyond the scope of this paper; further studies will be required to understand this issue in the quantitative manner.

Let us discuss astrophysical implication of our result. As described earlier, the polytropic exponent $\gamma$ is useful to describe a variety of components in the ISM. For example, the polytropic EOS with $\gamma$ $\simeq$ 0.8 can represent the density range of $10cm^{-3} \leq \ n \leq 10^4cm^{-3}$, where $n$ is hydrogen number density \citep{GM07b}. Giant molecular clouds can fall into this density range \citep{Fe01}. Also, the EOS with $\gamma$ $\sim$ 1.4 could represent the center of protostellar cores, which corresponds to the density range of $10^{12}cm^{-3} \leq \ n \leq 10^{17}cm^{-3}$ \citep{MI00}. Therefore, our result suggests that when driving of turbulence ceases to act, turbulence quickly decays in a corresponding dynamical timescale of a certain system irrespective of its spatial scale. Moreover, even if turbulence is initially driven by compressive driving, such as by supernova explosions, solenoidal motions will dominate as the turbulence decays due to much faster decay of compressive motions.

In summary, we have studied the influence of polytropic EOS and driving schemes on decaying turbulence and its statistics and found the following results.

\begin{enumerate} 
  	\item We have demonstrated that there is no significant correlation between scaling relation of the decay law (E $\propto$ $t^{-\alpha}$) and polytropic $\gamma$ in the case of solenodially driven turbulence.
    \item We have found that driving schemes have non-negligible effect on the decay rate of turbulence: the power-law index $\alpha$ for turbulence initially driven by compressive driving is smaller than that for turbulence initially driven by solenoidal driving.
    \item We have proven no significant effect of polytropic $\gamma$ on decay rate of velocity in compressively driven turbulence. 
    \item The polytropic $\gamma$ has small effect on the density fluctuations in compressively driven turbulence:  the smaller polytropic $\gamma$ is, the faster standard deviation of density fluctuation of the turbulence decays. 
    \item When we consider decay of solenoidal and compressive velocity components in compressively driven turbulence separately, energy of compressive velocity component decays much faster.
    \item Regarding statistics of compressively driven turbulence, we have shown deviation of the density PDF from a log-normal distribution, especially for $\gamma$ $>$ 1. In addition, we have found that skewness of the density PDF of the turbulence becomes zero as it decays. 
\end{enumerate}


\end{document}